\documentclass[aps,prev,twocolumn,preprintnumbers,floatfix,nofootinbib]{revtex4-1}
\pdfoutput=1

\newcommand{\be}{\begin{equation}}
\newcommand{\ee}{\end{equation}}
\newcommand{\bea}{\begin{eqnarray}}
\newcommand{\eea}{\end{eqnarray}}

\newcommand{\df}{\dfrac}

\newcommand{\sv}{\langle\sigma {\rm v}\rangle}

\usepackage{mathrsfs}
\usepackage{amsmath, amsthm, amssymb}
\usepackage{color}
\usepackage{epstopdf}
\usepackage[pdftex]{graphicx}
\usepackage{amssymb}
\usepackage{verbatim}
\usepackage{hyperref}
\usepackage{enumerate}

\begin{document}

\preprint{LPT--Orsay-13-36 }
\preprint{IFT-UAM/CSIC-13-67}
\preprint{ULB/TH-13-07}

\title{Thermal and non-thermal production of dark matter via  $Z'-$portal(s) }
\author{Xiaoyong Chu$^{a}$}
\email{xiaoyong.chu@ulb.ac.be}
\author{Yann Mambrini$^{b}$}
\email{yann.mambrini@th.u-psud.fr}
\author{J\'er\'emie Quevillon$^{b}$}
\email{jeremie.quevillon@th.u-psud.fr}
\author{Bryan Zald\'\i var$^{c}$}
\email{b.zaldivar.m@csic.es}

\vspace{1cm}
\affiliation{
${}^a$ Service de Physique Th\'eorique
Universit\'e Libre de Bruxelles, 1050 Brussels, Belgium \\
${}^b$ Laboratoire de Physique Th\'eorique 
Universit\'e Paris-Sud, F-91405 Orsay, France \\
${}^c$Instituto de Fisica Teorica, IFT-UAM/CSIC, 
 28049 Madrid, Spain 
}

\begin{abstract}
We study the genesis of dark matter in the primordial Universe for representative classes of $Z'$-portals models.
 For weak-scale $Z'$ mediators we compute the range of values of the kinetic mixing allowed by WMAP/PLANCK
 experiments corresponding to a FIMP regime. We show that very small values of $\delta$ ($10^{-12} \lesssim \delta \lesssim 10^{-11}$)
 are sufficient to produce the right amount of dark matter.
  We also analyse the case of very massive gauge mediators, whose mass $m_{Z'}$ is larger than the reheating temperature, $T_{\mathrm{RH}}$, with a weak--scale coupling $g_D$ to 
  ordinary matter. 
  Relic abundance constraints then impose a direct correlation between $T_{\mathrm{RH}}$
   and the effective scale $\Lambda$ of the interactions:
  $\Lambda \sim 10^3-10^5 \times T_{\mathrm{RH}}$.
  Finally we describe in some detail the process of dark thermalisation and study its consequences on the computation
  of the relic abundance.

\end{abstract}
\maketitle
\section{Introduction}

Even if PLANCK \cite{Ade:2013zuv} confirmed recently the presence of Dark Matter (DM) in the Universe with an unprecedented precision, its
nature and its genesis are still unclear. 
The most popular scenario for the DM evolution is based on the mechanism of ``thermal freeze-out" (FO)
\cite{Bergstrom:2000pn,Bertone:2004pz}. 
In this scenario DM particles $\chi$ are initially in thermal equilibrium with respect to the thermal bath. When the temperature of 
the hot plasma $T$ in the early Universe dropped below the DM mass, its population decreased exponentially until the annihilation rate
 into lighter species $\Gamma_\chi$ could not overcome the expansion rate of the Universe driven by the Hubble parameter $H(T)$.
  This defines the freeze-out temperature: $H(T_{\mathrm{FO}})\gtrsim \Gamma_\chi$.
 The comoving number density  of the DM particles\footnote{Proportional to the yield $Y_\chi = n_\chi/{\bf s}$,  $n_\chi$ being the physical density
  of dark matter particles and ${\bf s}$ the entropy density.} and thus its relic abundance
  are then fixed to the value
that PLANCK \cite{Ade:2013zuv} and WMAP\cite{WMAP} observe today, $\Omega h^2 =0.1199 \pm 0.0027$ at 68\% CL. 
In this scenario it is obvious that the stronger the interaction between DM and the rest of the thermal bath is, 
the more DM pairs annihilate, ending-up with smaller relic densities. 
 The detection prospects for frozen-out WIMPs are remarkable, since they involve cross-sections which can be
 probed nowadays with different experimental strategies, as production at colliders\cite{Dreiner:2013vla}, Direct Detection (DD) and Indirect Detection (ID) experiments \cite{Cheung:2012gi}.

 This popular freeze-out scenario is based on the hypothesis that the dark matter is initially produced at a democratic rate 
 with the Standard Model (SM) particles. The so-called ``WIMP miracle" can then be obtained if dark matter candidate has a mass of the electroweak scale and the dark sector and the Standard Model sector interact through electroweak strength coupling. Alternatively one can relax the hypothesis of democratic production rate and suppose that
 the initial abundance of dark matter has been negligibly small whether by hierarchical or gravitational coupling to the inflaton or others mechanisms.
 This is the case for gravitino DM  \cite{Moroi:1993mb}, Feebly Interacting Massive Particle dark matter (FIMP) 
 in generic scenarios \cite{Hall:2009bx,Yaguna:2011ei,
 Chu:2011be}, scalar portals \cite{McDonald:2001vt,Yaguna:2011qn}, decaying dark matter \cite{Arcadi:2013aba} 
 or Non Equilibrium Thermal Dark Matter (NETDM) \cite{Mambrini:2013iaa}.

Alternatively to the freeze--out, in the freeze-in (FI) mechanism the DM gets populated through interactions and decays from particles of the thermal bath with such an extremely weak rate 
(that is why called FIMP) that it never reaches thermal equilibrium with the plasma. In this case, the dark matter population $n_\chi$ grows
 very slowly until the temperature of the Universe 
 %$T_{\mathrm{FI}}$
drops below the mass $m_\chi$. The production mechanism is then frozen by the expansion rate of the Universe $H(T_{\mathrm{FI}})$. 
Contrary to the
 FO, in the FI scenario the stronger the interaction is, the larger the relic density results at the end, provided that the process never thermalises with the thermal bath. 
 Due to the smallness of its coupling, the dark matter becomes very difficult to detect in colliders or direct detection experiments.  However, one of the predictions of this scenario
  is that (visible) particles possibly decaying to dark matter need to have a long lifetime\cite{Hall:2009bx}, so this peculiarity can be probed in principle in the LHC
  for example through the analysis of displaced vertices.   

Very recently, it was analysed in \cite{Mambrini:2013iaa} a scenario where the dark matter was also produced out-of-equilibrium, but differing from the orthodox
 FI mechanism in an essential way. In this new NETDM proposal the DM-SM couplings can be large (as for FO case),
  whereas the particle mediating the interaction is very heavy, which caused the evolution of dark matter number density to be dominated mostly by very high 
  temperatures, just after the reheating epoch. This situation is opposite to the FI scenario where the couplings are feeble, typically ${\cal O}(10^{-11})$, 
  and the portal is either massless or at least has a mass smaller than dark matter mass $m_\chi$, causing the process to be dominated by low temperatures
   $(T\lesssim m_{\chi})$ instead.

In this work we study the dark matter candidate $\chi$ populated by vector-like portals, whose masses lie in two different regimes: 1) A very heavy 
mediator, through the study of effective interactions of dark matter with the SM\footnote{Note that in this analysis, the nature of the mediator (vector or scalar) is 
not fundamental and our result can apply for the exchange of heavy scalars or heavy Higgses present in unified models also.},
 and 2) An intermediate mediator, through the analysis of 
a kinetic-mixing model which contains a $Z'$ acting as the portal. This study complements the case 
of massless vector-like mediators, studied in \cite{Chu:2011be}, showing distinct features concerning the evolution of the dark-sector
 independent thermalisation. On the other hand, we show the characteristics of the NETDM mechanism for a general vector-like interaction.
\newline\newline
The paper is organised as follows. In section II a brief summary of non-thermalised production of dark matter particles is presented. Section III is devoted
 to present the two models of study, whose results are described in detail in section IV, before concluding in section V. 

\section{Boltzmann equation and production of Dark Matter out of equilibrium}

\noindent
If we consider that in the early stage of the Universe the abundance of dark matter has been
negligibly small whether by inflation or some other mechanism,
the solution of the Boltzmann equation
 can be solved numerically in effective cases  like in \cite{Hall:2009bx} or in the case of the exchange 
 of a massless hidden photon as did the authors of \cite{Chu:2011be}.
 Such an alternative to the classical freeze out thermal scenario was in fact proposed earlier
   in \cite{McDonald:2001vt} in the framework of the Higgs-portal
 model \cite{Yaguna:2011qn} and denominated ``freeze in"  \cite{Hall:2009bx}. If one considers a massive field $Z'$ coupling to the dark matter,
the dominant processes populating 
the DM particle $\chi$ are given by the decay  $Z'\to \bar\chi\chi$ and the annihilation
 $\overline{SM} ~ SM \to \bar\chi\chi$ involving 
the massive particle $Z'$  as a mediator, or ``portal" between the visible (SM) sector and the invisible (DM) sector.  Our study will
be as generic as possible by taking into account both processes at the same time, although we will show that 
for very large mediator masses $m_{Z'}$, or if the $Z'$ is not part of the thermal bath, the decay process is highly suppressed, and the annihilation clearly 
dominates\footnote{Note that in \cite{Hall:2009bx} 
the 2$\to$2 annihilation process is considered subdominant with respect to the 1$\to$2 decay process. 
However in the scenarios we will study, the annihilation dominates.}. 
Under the Maxwell--Boltzmann approximation\footnote{We have checked that the Maxwell-Boltzmann approximation 
induces a 10\% error in the solution which justifies it to understand the general result. See \cite{bfmz} for an explicit cross-check of this approximation.} 
 one can obtain an analytical solution of the DM yield adding the annihilation and decay processes:

\bea
\label{Boltz1}
Y_\chi &\approx& \left[\left(\frac{45}{\pi}\right)^{3/2}\frac{M_{\rm p}}{4\pi^2}\right] \int_{T_0}^{T_{\rm RH}}dT\int_{4m_\chi^2}^{\infty}ds \df{1}{\sqrt{g_*}g_*^s}\df{1}{T^5} \nonumber
\\
&\times&  K_1\left(\df{\sqrt{s}}{T}\right) \df{1}{2048\pi^6}\sqrt{s-4m_\chi^2}|\tilde{\cal M}_{2\to2}|^2~ \nonumber  \\
&+& \left[\left(\frac{45}{\pi}\right)^{3/2}\frac{M_{\rm p}}{4\pi^2}\right] \int_{T_0}^{T_{\rm RH}}dT \df{1}{\sqrt{g_*}g_*^s}\df{1}{T^5} \label{FI:Production} \\
&\times& K_1\left(\df{m_{Z'}}{T}\right) \df{1}{128\pi^4}\sqrt{m_{Z'}^2-4m_\chi^2}|\tilde{\cal M}_{1\to2}|^2~,\nonumber 
\eea

\noindent
where $M_{\rm p}$ is the Planck mass, $T_0=2.7$ K the present temperature of the Universe, $T_{\rm RH}$ the reheating temperature,
and $K_1$ is the 1st-order modified Bessel function of the second kind, $g_*, g_*^s$ are the effective numbers of degrees of freedom of the thermal bath for the energy and entropy densities respectively.  Finally, $|\tilde{\cal M}_{i\to2}|^2\equiv \int d\Omega |{\cal M}_{i\to2}|^2$, where ${\cal M}_{i\to2}$ is the squared amplitude of the
   process $i\to2$ summed over all initial and final degrees of freedom, and $\Omega$ is the standard solid angle.
   Then, assuming a symmetric scenario for which the populations of $\chi$ and $\bar\chi$ are produced at the same rate,
    we can calculate the relic density
\be
\Omega_\chi h^2 \approx \df{m_\chi Y^0_\chi}{3.6\times10^{-9}{\rm GeV}}~,
\label{Om0}
\ee
where the super-index ``0" refers to the value measured today. 
%Note that in (\ref{Boltz1}) we have considered the reheating temperature $T_{\rm RH}$ as the cut-off for the integral over $T$. A 
%priori, from cosmological arguments, it doesn't make sense to go beyond $T_{\rm RH}$, since this is precisely the epoch in 
%which the thermal bath starts to play a role. 
It turns out that the yield of the DM is actually sensitive to the temperature at which the
DM is largely produced: at the beginning of the thermal history of the Universe if the mediator mass lies 
above the reheating temperature $m_{Z'} > T_{\mathrm{RH}}$ (the so--called NETDM scenario
 \cite{Mambrini:2013iaa}), or around the mass of the mediator if $2m_\chi< m_{Z'} < T_{RH}$
 as the Universe plasma reaches the pole of the exchanged particle, in a resonance--like effect.
 Note that in the case of massless hidden photon or effective freeze--in cases described respectively in \cite{Chu:2011be} and \cite{Hall:2009bx}
 the effective
 temperature scale defining the nowadays relic abundance
  is given by the only dark scale accessible, i.e. the mass of the DM (like in the classical freeze out scenario).
 %On the other hand, the results are completely insensible to the lower bound $T_0$, because for temperatures $T<<m_\chi$ the Boltzmann suppression in huge, meaning %that there is no enough 
% energy in the plasma to produce DM particles. 
In the following sections we will describe the two microscopic frameworks ($m_{Z'} > T_{\mathrm{RH}}$ and $ m_{Z'} < T_{\mathrm{RH}}$) in which 
we have done our analysis.

%======================================================================================
%===============================  MODELS =============================================
%======================================================================================

\section{The models}

%=====================================   MODEL MZ' > TRH   ================================================

\subsection{$m_{Z'} > T_{\mathrm{RH}}$ : effective vector-like interactions}

\noindent
If interactions between DM and SM particles involve very heavy 
 particles with masses above the reheating temperature $T_{RH}$, we can describe them in the framework of effective field theory as a Fermi--like interaction
 can be a relatively accurate description of electroweak theories when energies involved
 in the interactions are below the electroweak scale.
 %This is typically the case if the mediator exchanged between the dark sector (DM) and the visible world (SM) has a mass larger than the maximum energy present in the thermal history of the Universe, the reheating temperature $T_{\rm{RH}}$.
 Several works studying effective interactions in very different contexts have been done by the authors of
\cite{Bai:2010hh}-\cite{Chae:2012bq}  for accelerator constraints and \cite{Gao:2011ka}-\cite{Gondolo:2012vh} for some DM aspects.
Depending on the nature of the DM we will consider the following effective operators, for complex scalar
and Dirac fermionic DM \footnote{Other operators of the $\gamma_\mu \gamma^5$ pseudo-scalar types for instance can also
appear for chiral fermionic DM, but we will neglect them as  they bring similar contribution to the annihilation process.}:

\vspace{0.5cm}

\noindent
\underline{Fermionic dark matter}:
\bea
&&
{\cal O}_V^f =\df{1}{\Lambda_f^2}(\bar f\gamma^\mu f)(\bar\chi\gamma_\mu\chi)~,
\eea
leading to the squared-amplitude:
\bea
&&
|{\cal M}_V^f|^2 = \df{32 N^f_c}{\Lambda_f^4} \left\{\df{s^2}{8}+2\left(\df{s}{4}-m_f^2\right)\left(\df{s}{4}-m_\chi^2\right)\cos^2\theta \right.
\nonumber
 \\
&&
+\left. \df{s}{2}(m_\chi^2+ m_f^2)\right\}~.
\label{OVf}
\eea

\underline{Scalar dark matter}:
\bea
&&
{\cal O}_V^s = \df{1}{\Lambda_f^2} (\bar f\gamma^\mu f)[(\partial_\mu\phi)\phi^* - \phi(\partial_\mu \phi)^*]
\eea
which leads to:
\bea
&&
|{\cal M}_V^s|^2 = 4\df{N_c^f}{\Lambda_f^4} \left[-8 \left(\df{s}{4}-m_f^2\right)\left(\df{s}{4}-m_\phi^2\right)\cos^2\theta \right. 
\nonumber
 \\
&&
+ \left. \left(\df{s}{2}-m_f^2\right)(s-4m_\phi^2) + m_f^2(s-4m_\phi^2)\right].
\label{OVs}
\eea

\noindent
As we will show in section \ref{sec:Res-Eff}, the main contribution to the population of DM in this case occurs around the reheating
time. At this epoch, all SM particles $f$ and the DM candidate  $\chi$
can be considered as massless relativistic species.\footnote{This is justified numerically by the fact that large $s$ (\,$\gtrsim 4T^2$\,$\gg m^2_\chi(T), m^2_f(T)$\,) dominates the first integration in Eq.(\ref{FI:Production}).} The expressions (\ref{OVf}, \ref{OVs}) then become

\bea
\label{Eq:effectiveapprox}
&&
|{\cal M}_V^f|^2 \approx 4\frac{N^f_c}{\Lambda_f^4} ~ s^2(1 +  \cos^2 \theta), \nonumber \\
&&
|{\cal M}_V^s|^2 \approx 2\frac{N_c^f}{\Lambda_f^4} ~s^2(1 -  \cos^2 \theta),
\eea

\noindent
where, for simplicity and without loss of generality,
we have considered universal effective scale $\Lambda_f \equiv \Lambda$. 
Considering different scales in the hadronic and leptonic sectors as was done 
in \cite{Mambrini:2011pw} for instance won't change appreciably our conclusions.
 %as this effective dependance can be absorbed in the degrees of freedom of the annihilating particles.

%===============================   MODEL MZ' < TRH  ==========================================

\subsection{$m_{Z'}< T_{\mathrm{RH}}$: extra $Z'$ and kinetic mixing}

\subsubsection{Definition of the model}

Neutral gauge sectors with an additional dark $U'(1)$
symmetry in addition
to the SM hypercharge $U(1)_Y$ and an associated $Z'$
are among the best motivated extensions of the SM, and give the possibility
that a DM candidate lies within this new gauge sector of the theory.
Extra gauge symmetries are predicted in most Grand Unified Theories (GUTs)
and appear systematically in string constructions. Larger groups than $SU(5)$
or $SO(10)$ allow the SM gauge group and $U'(1)$ to be embedded into
bigger GUT groups.
Brane--world $U'(1)$s are special compared to GUT $U'(1)$'s because there
is no reason for the SM particles to be charged under them;
for a review of the phenomenology of the extra $U'(1)$s generated in such
scenarios see e.g.
 \cite{Langacker:2008yv}.
In such framework, the extra $Z'$ gauge boson would act as a portal
between the ``dark world''
(particles not charged under the SM gauge group) and the ``visible'' sector.

\noindent
Several papers considered that the ``key" of the portal could be the gauge
invariant kinetic mixing $(\delta/2) F_Y^{\mu \nu} F'_{\mu \nu}$ \cite{Holdom, Dienes:1996zr}.
One of the first models of DM from the hidden sector with a
massive additional $U'(1)$, mixing with the SM hypercharge through both mass and kinetic
mixings,
can be found in \cite{Feldman:2006wd}.
The DM candidate $\chi$ could be the lightest (and thus stable) particle of
this secluded sector. Such a mixing has been justified in
recent string constructions \cite{Cicoli:2011yh,Kumar:2007zza,Goodsell:2011wn,
Javier,Cassel:2009pu},  supersymmetry \cite{Andreas:2011in},
SO(10) framework \cite{Krauss:2013jva} but has also been studied within a model independent
approach \cite{Feldman:2007wj,Pospelov:2008zw,Mambrini:2010yp}
with vectorial dark matter \cite{Domingo:2013tna} or extended extra-$U(1)$ sector
\cite{Heeck:2011md}. For typical smoking gun signals in such models, like a monochromatic gamma-ray line, see 
\cite{monogamma}.

The matter content of any dark $U'(1)$ extension of the SM can be decomposed
into three families of particles:

\begin{itemize}
\item{The $Visible$ $sector$ is made of particles which are charged under the SM
gauge group $SU(3)\times SU(2)\times U_Y(1)$ but not charged under $U'(1)$
(hence the ``dark'' denomination for this gauge group)}.
\item{The $Dark$ $sector$ is composed of the particles charged under
$U'(1)$ but neutral with respect to the SM gauge symmetries. The DM
($\chi$) candidate is the lightest particle of the dark sector}.
\item{The $Hybrid$ $sector$ contains states with SM $and$ $U'(1)$ quantum numbers. 
These states are fundamental because they act as a portal between
the two previous sectors through the kinetic mixing they induce at loop
order.} 
\end{itemize}

\noindent
From these considerations, it is easy to build the effective Lagrangian
generated at one loop :
\begin{eqnarray}
{\cal L}&=&{\cal L}_{\rm{SM}}
-\frac{1}{4} \tilde B_{\mu \nu} \tilde B^{\mu \nu}
-\frac{1}{4} \tilde X_{\mu \nu} \tilde X^{\mu \nu}
-\frac{\delta}{2} \tilde B_{\mu \nu} \tilde X^{\mu \nu}
\nonumber
\\
&+&i\sum_i \bar \psi_i \gamma^\mu D_\mu \psi_i
+i\sum_j \bar \Psi_j \gamma^\mu D_\mu \Psi_j\,,
\label{Kinetic}
\end{eqnarray}
$\tilde B_{\mu}$ being the gauge field for the hypercharge, 
$\tilde X_{\mu}$ the gauge field of $U'(1)$ and
$\psi_i$ the particles from the hidden sector, $\Psi_j$ the particles
 from the hybrid sector, 
$D_{\mu}  =\partial_\mu -i (q_Y \tilde g_Y \tilde B_{\mu} + q_D \tilde g_D
 \tilde X_{\mu} + g T^a W^a_{\mu})$, $T^a$ being the $SU(2)$ generators, and 
\begin{equation}
\delta= \frac{\tilde g_Y \tilde g_D}{16 \pi^2}\sum_j q_Y^j q_D^j 
\log \left( \frac{m_j^2}{M_j^2} \right)
\end{equation}
with $m_j$ and $M_j$ being hybrid mass states \cite{Baumgart:2009tn}\,. It has been showed \cite{Dienes:1996zr} that the value of $\delta $ may be as low as $10^{-14}$, e.g. in the case of gauge-mediated SUSY-breaking models, where the typical relative mass splitting ${|M_j -m_j| / M_j }$ is extremely small. 

%with $\mu$ being the mass scale at which you make the experiment 
%($\sim m_{\psi_0}$ in our case) and $q^j_Y$ ($q^j_D$) the $U(1)_Y$ ($U'(1)$)
 %charges. It is worth noticing that in consistent constructions, 
 %the dependance on the scale $\mu$ disappears and the argument of the log depends
 %only on the ratio of hybrid states masses $M'/M$\cite{Baumgart:2009tn} .
\noindent
Notice that the sum is on all the hybrid states, as they are the only ones which can contribute to the
 $\tilde B_{\mu},\, \tilde X_{\mu}$ propagator.
After diagonalising of the current eigenstates that makes the gauge kinetic
terms of Eq.(\ref{Kinetic}) diagonal and canonical, 
we can write after the $SU(2)_L\times U(1)_Y$ breaking\footnote{Our notation
for the gauge fields are 
($\tilde B^\mu,\tilde X^\mu$) before the diagonalization, 
($B^\mu, X^\mu$) after diagonalization and 
($Z^\mu,Z'^\mu$) after the electroweak breaking.}
\begin{eqnarray}
A_{\mu} &=& \sin \theta_W W_{\mu}^3 + \cos \theta_W B_{\mu}
\\
Z_{\mu} &=& \cos \phi ( \cos \theta_W W_{\mu}^3 - \sin \theta_W B_{\mu})
- \sin \phi  X_\mu
\nonumber
\\
Z'_{\mu}&=&\sin \phi (\cos \theta_W W_\mu^3 - \sin \theta_W B_\mu)
+ \cos \phi  X_\mu
\nonumber
\end{eqnarray}
with, to first order in $\delta$,
\begin{eqnarray}
\cos \phi &=& \frac{\alpha}{\sqrt{\alpha^2 + 4 \delta^2 \sin^2 \theta_W}}
~~
\sin \phi = \frac{2 \delta \sin \theta_W}{\sqrt{\alpha^2 + 4 \delta^2 \sin^2 \theta_W}}
\nonumber
\\
\alpha &=& 1- m^2_{Z'}/M^2_Z - \delta^2 \sin^2 \theta_W
\label{sphi}
\\
&\pm& \sqrt{(1-m^2_{Z'}/M^2_Z -\delta^2 \sin^2 \theta_W)^2+ 4 \delta^2 \sin^2 \theta_W}
\nonumber
\end{eqnarray}
\noindent
and + (-) sign if $m_{Z'}< (>)M_Z$.
The kinetic mixing parameter $\delta$ generates an effective coupling of 
SM states $\psi_{\rm{SM}}$ to $Z'$, and a coupling of $\chi$ to 
the SM $Z$ boson which
induces an interaction on nucleons.
Developing the covariant derivative on SM and $\chi$ fermions state,
we computed the effective $\psi_{\rm{SM}}\psi_{\rm{SM}}Z'$  and
 $\chi\chi Z$ couplings to first order\footnote{  One can find
  a detailed analysis of the spectrum and couplings of the model
   in the appendix of Ref.\cite{Mambrini:2010yp}. The coupling $g_D$ is the effective dark coupling $\tilde g_D$
   after diagonalization.}
    in $\delta$ and obtained
   \bea
   {\cal L}= q_D \tilde g_D (\cos \phi~ Z'_\mu \bar \chi \gamma^{\mu} \chi +
   \sin \phi~ Z_\mu \bar \chi \gamma^{\mu} \chi ).
   \eea
   
   \noindent
   In the rest of the analysis, we will use the notation $\tilde g_D \rightarrow g_D$. 
 We took $q_D g_D=1$ through our analysis, keeping in mind that for the $m_{Z'}$-regimes we consider here, our results stay 
 completely general by a simple rescaling of the kinetic mixing $\delta$  if the dominant process transferring energy from SM to DM is $\bar ff\rightarrow Z^{'(*)}\rightarrow \bar \chi\chi$; whereas if processes involving on-shell $Z'$ dominate, the results become nearly independent of $q_Dg_D$.

\subsubsection{Processes of interest}

As is clear from the model defined above, both DM and SM particles will interact via the standard $Z$ or the extra $Z'$ boson. Thus a priori there are four processes contributing to the DM relic abundance: $\bar ff \to V\to\bar\chi\chi$, and  $V\to \bar\chi\chi$, where $V$ can be $Z$ and/or $Z'$, and in the 2 $\to2$ process both $Z$ and $Z'$ interfere to produce the total cross-section.\footnote{There are additional processes, not written here, which can have non-negligible influence on the final DM number density; e.g. $\bar ff\to ZZ'\to Z \bar\chi\chi$, with a $t$-channel exchange of a fermion $f$. These processes have been taken into account in the full numerical solution of the coupled set of Boltzmann equations, as shown below.} The amplitudes of those processes are:

\be
\label{procKM1}
|{\cal M}_{2\to2}|^2 = |{\cal M}_{Z}|^2 + |{\cal M}_{Z'}|^2 + ({\cal M}_{Z}{\cal M}^*_{Z'} + {\rm h.c.})~,
\ee
where
\bea
\label{procKM2}
|{\cal M}_{Z}|^2 &=& \df{(q_Dg_D)^2\sin^2\phi}{(s-M_Z^2)^2+(M_Z\Gamma_Z)^2} \\
&\times& \left\{ (c_L^2+c_R^2)\left[16 m_\chi^2m_f^2(\cos^2\theta-\sin^2\theta)\right.\right. \nonumber \\
&+&8m_\chi^2s\sin^2\theta-\left.8m_f^2s\cos^2\theta+2s^2(1+\cos^2\theta) \right] \nonumber \\
&+&\left. c_L c_R (32m_\chi^2m_f^2 + 16 m_f^2 s) \right\}~, \nonumber
\eea

\bea
\label{procKM3}
&~&|{\cal M}_{Z'}|^2 = |{\cal M}_{Z}|^2  ~~~{\rm with:} ~[\sin\phi\to\cos\phi, \\
&~&~ (M_Z,\Gamma_Z)\to(m_{Z'},\Gamma_{Z'}), ~~(c_L,c_R)\to(c'_L,c'_R)]~, \nonumber
\eea
and
\bea
&&{\cal M}_{Z}{\cal M}^*_{Z'} + {\rm h.c.}= \df{2A~(q_Dg_D)^2\sin\phi\cos\phi}{A^2+B^2}  \nonumber\\
&&\times\left\{ (c_L c'_L+c_R c'_R)\left[16 m_\chi^2m_f^2(\cos^2\theta-\sin^2\theta)\right.\right. \nonumber \\
&&+8m_\chi^2s\sin^2\theta-\left.8m_f^2s\cos^2\theta+2s^2(1+\cos^2\theta) \right] \nonumber \\
&&+\left. (c_L c'_R + c_R c'_L) (16m_\chi^2m_f^2 + 8 m_f^2 s) \right\}~,
\label{procKM4}
\eea
with
\bea
&& A = s^2-s(M_Z^2+m_{Z'}^2) + M_Z^2 m_{Z'}^2 + M_Z m_{Z'}\Gamma_Z\Gamma_{Z'} \nonumber \\
&& B = s(\Gamma_Z M_Z - \Gamma_{Z'}m_{Z'}) + M_Z^2 m_{Z'}\Gamma_{Z'} - m_{Z'}^2 M_Z \Gamma_Z~, \nonumber\\
&&
\eea
whereas for the 1 $\to$ 2 process we have:

\be
 |{\cal M}_{1\to2}|^2 = \left\{ \begin{array}{ll}
4(q_D g_D)^2(\sin^2\phi) (M_Z^2+2m_\chi^2) & {\rm if} ~~V = Z\\
4(q_D g_D)^2(\cos^2\phi) (m_{Z'}^2+2m_\chi^2) & {\rm if} ~~V = Z' ~.\\
\end{array} \right.
\label{procKM5}
\ee
Here the coefficients $c_{L,R}$ and $c'_{L,R}$ are the left and right couplings of the SM fermions to the $Z$ and $Z'$ bosons, respectively.
 Their explicit forms are shown in the appendix.

%======================================================================================
%===============================  RESULTS  =============================================
%======================================================================================

\section{Results and discussion}

%=====================================   RESULTS MZ' > TRH   ================================================

\subsection{$m_{Z'} > T_{\rm{RH}}$}
\label{sec:Res-Eff}

\begin{figure}
% $T_{RH}$\includegraphics[width=0.5\textwidth,angle=0]{Reheat.pdf}
%\includegraphics[width=0.5\textwidth,angle=0]{YvsX_vecscalDM.pdf}
\includegraphics[width=0.5\textwidth,angle=0]{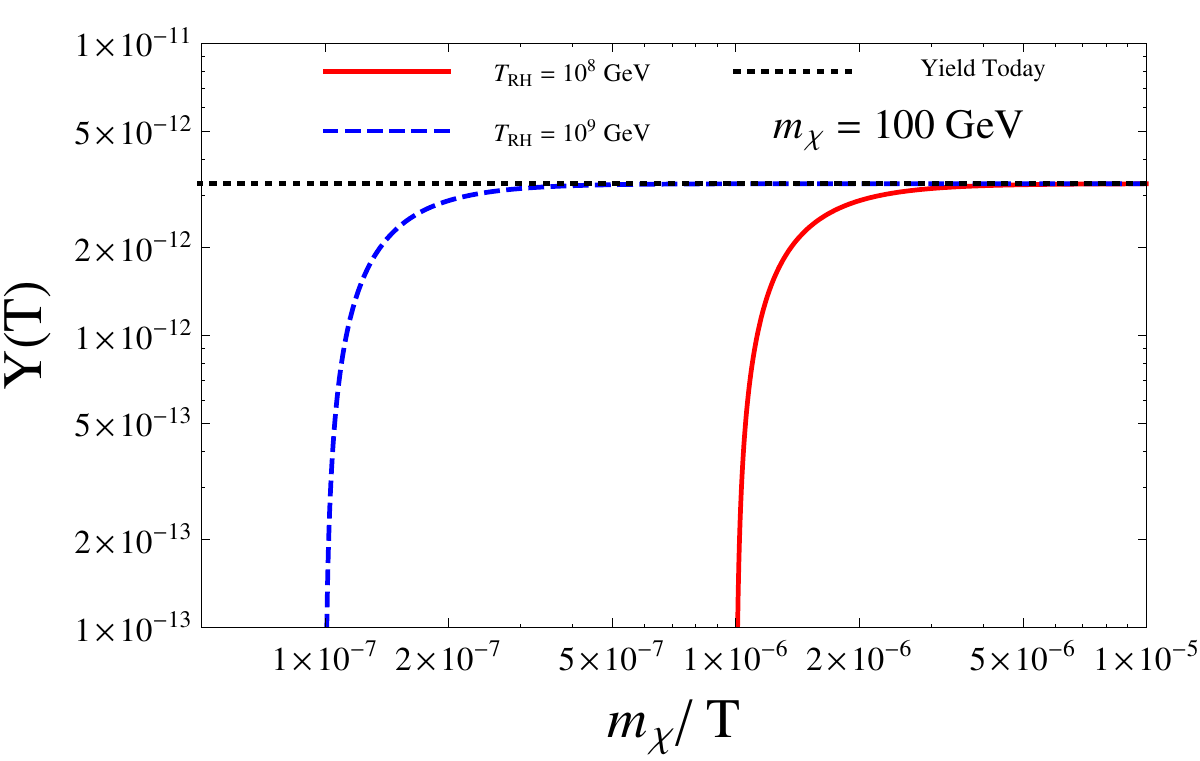}
\caption{\footnotesize{Evolution of the number density per comoving frame ($Y = n/{\bf s}$) for a 100 GeV fermionic DM
 as a function
of $m_\chi/T$ for two reheating temperatures, $T_{RH}= 10^8$ (red) and $10^9$ (blue) GeV in the case of vector
interaction for fermionic a DM candidate. The value of the scale $\Lambda$
has been chosen such that the nowadays yield $Y$
corresponds to the nowadays value of $Y(T_0)$ measured by WMAP: $Y(T_0) \simeq 3.3 \times 10^{-12}$
represented by the horizontal black dashed line (see the text for details). }}
\label{Fig:Y(T)}
\end{figure}

\noindent
In the case of production of DM through SM particle annihilation,
 the Boltzmann equation can be simplified

\bea
\frac{dY}{dx} &=& \frac{1}{16(2 \pi)^8} \frac{1}{g_* \sqrt{g_*^s}} \left( \frac{45}{\pi} \right)^{3/2} \frac{M_p}{m_\chi} 
\label{Boltzmann}
\\
& \times &
\int_{2x}^\infty z \left(z^2 - 4 x^2 \right)^{1/2} K_1(z) dz |{\cal M}(z)|^2 d\Omega
\nonumber
\eea 

\noindent
with $z = \sqrt{s}/T$, $x = m_\chi/T$ and $\Omega$ the solid angle of the outgoing DM particles.
Using the expression for $|{\cal M }|^2$ obtained in Eq.(\ref{Eq:effectiveapprox}) 
we can write an analytical expression of the relic yields present nowadays if we suppose
(as we will check) that the non-thermal production of DM happens at temperatures
(and thus $s$) much larger than the mass of DM or SM particles
($m_f, m_\chi \ll \sqrt{s}$). After integrating over the temperature ($x$ to be precise) from $T_{RH}$ to $T$,
and considering that
all the fermions of the SM contribute democratically ($\Lambda_f \equiv \Lambda$) one obtains\footnote{Notice
that the factor of difference corresponds to the different degrees of freedom for a real scalar and Dirac fermionic DM.}

\bea
&&
Y_V^f (T) \simeq \frac{4}{3}\frac{384}{(2 \pi)^7} \left( \frac{45}{\pi g_*^s}\right)^{3/2} \frac{M_p}{\Lambda^4}
\biggl[ T_{RH}^3 - T^3 \biggr],
\nonumber
\\
&&
Y_V^s(T) \simeq \frac{1}{3}\frac{384}{(2 \pi)^7} \left( \frac{45}{\pi g_*^s}\right)^{3/2} \frac{M_p}{\Lambda^4}
\biggl[ T_{RH}^3 - T^3 \biggr],
\label{Eq:Y(T)}
\eea

\noindent
where $g_*\sim g_*^s$ has been used. We show in Fig.~(\ref{Fig:Y(T)}) the evolution of $Y(T)$ for a fermionic DM as a function of $x=m_\chi/T$ with $m_\chi=100$ GeV for two different reheating temperatures, $T_{\rm{RH}}=10^8$ and $10^9$ GeV. We note that to obtain analytical
solution to the Boltzmann equation, we approximated the 
Fermi-Dirac/Bose-Einstein by Maxwell-Boltzmann distribution. This can introduce a 10\% difference with respect to the exact 
case \cite{bfmz}. However, when performing our study we obviously solved numerically the complete set of Boltzmann 
equations.  As one can observe in Fig.~(\ref{Fig:Y(T)}),
 the relic abundance of the DM is saturated very early in the Universe history, around $T \simeq T_{\rm{RH}}$,
 confirming our hypothesis that we can consider all the particles in the thermal bath (as well as the DM)
 as massless in the annihilation process: $m_\chi, m_f \ll \sqrt{s}$. At $T \simeq T_{RH}/2$ the DM already reaches its
asymptotical value.

%%%%%%%%%%%%%%%%%%%%%%%%%%%%%%%%%%%%%%%%%%%%%%%%%%%%%%%%%%%%%%%%%%%
%%%%%%%%%%%%%%%%%%%%%%%%%%%%%%%%%%%%%%%%%%%%%%%%%%%%%%%%%%%%%%%%%%%%%

\noindent
Moreover, for a given value of the reheating temperature $T_{\rm{RH}}$, 
we compute the effective scale $\Lambda$  such that the present DM yield $Y(T_0)$
respects the value measured by WMAP/PLANCK :  $Y(T_0) \simeq 3.3 \times 10^{-12}$ for a 100 GeV DM.
Imposing this constraint in Eq.(\ref{Eq:Y(T)}), we obtain $\Lambda({T_{RH} = 10^8 \mathrm{GeV}})\simeq 3.9 \times 10^{12}$ GeV  
 and $\Lambda({T_{RH} = 10^9 \mathrm{GeV}})\simeq 2.2 \times 10^{13}$ GeV for a fermionic  DM.

%\bea
%&&
%(\Lambda_V^f)^4 =  \frac{4}{3} \frac{384}{(2 \pi)^7} \left( \frac{45}{\pi g_s}\right)^{3/2} \frac{M_p T_{RH}^3}{3.3 \times 10^{-12}}
%\left( \frac{m}{100~ \mathrm{GeV}} \right) \left( \frac{0.1}{\Omega h^2} \right)
%\nonumber
%\\
%&&
%(\Lambda_V^s)^2 =  \frac{1}{3} \frac{24}{(2 \pi)^7} \left( \frac{45}{\pi g_s}\right)^{3/2} \frac{M_p T_{RH}}{3.3 \times 10^{-12}}
%\left( \frac{m}{100~ \mathrm{GeV}} \right) \left( \frac{0.1}{\Omega h^2} \right)~. \nonumber\\
%&&~
%\label{Eq:lambda}
%\eea

\begin{figure}[ht]
\centering
\includegraphics[width=0.5\textwidth,angle=0]{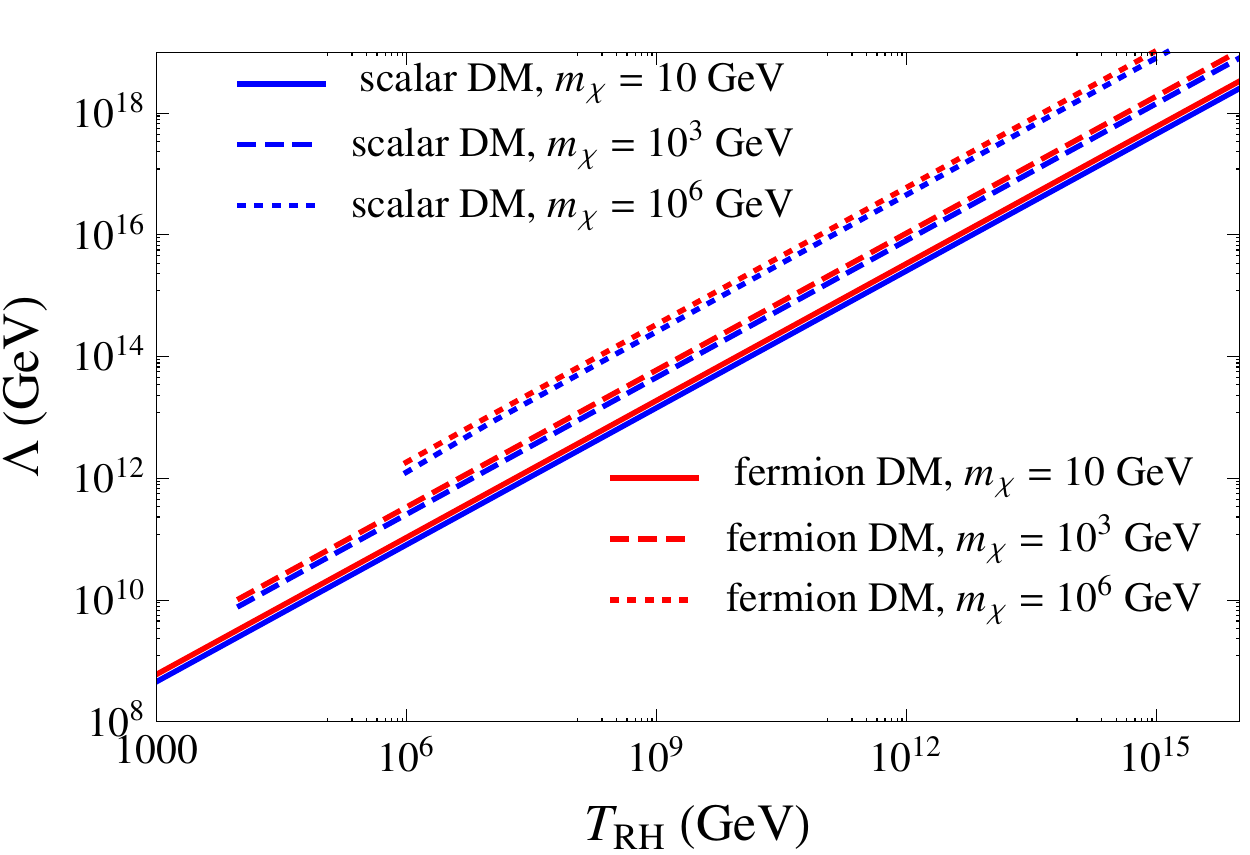}
\caption{\footnotesize{Values of the scale $\Lambda$ for fermionic (red) and scalar (blue) DM, 
assuming good relic abundance ($\Omega_{\chi}h^2 = 0.12$) and DM mass of 10 GeV (solid), 1000 GeV (dashed) and $10^6$ GeV (dotted),
 as a function of the reheating temperature.}}
\label{fig:EffOps2}
\end{figure}

\noindent
As a consequence, we can derive
the value of $\Lambda$ respecting the WMAP/PLANCK constraint as a function of the reheating temperature $T_{\rm{RH}}$ for different masses of DM. 
This is illustrated in Fig.~(\ref{fig:EffOps2}) where we solved numerically the exact Boltzmann
equation. We observe that the values of $\Lambda$ we obtained with our analytical solutions -extracted from Eqs.(\ref{Eq:Y(T)})- are pretty accurate
and the dependence on the nature  (fermion or scalar) of the DM is very weak.
We also  notice that the effective scale needed to respect WMAP constraint is very consistent 
with GUT--like SO(10) models which predict typical $10^{12-14}$ GeV as intermediate scale if one imposes unification
\cite{Mambrini:2013iaa}. 
%The corresponding scale for a scalar dark matter in scalar--type interaction is obviously larger than in the fermionic case due to the reduced suppression ($1/\Lambda_S^s$ vs $1/(\Lambda_i^j)^2$) of the coupling.  
Another interesting point is that $\Lambda \gg T_{RH}$ whatever is the nature of DM, ensuring
the coherence of the effective approach. 
We have also plotted the result for very heavy DM candidates (PeV scale) to show that in such a scenario, there is no need
for the DM mass to lie within electroweak limits, avoiding any ``mass fine tuning" as in the classical WIMP paradigm.

%Indeed,  for some reasons $T_{RH} \lesssim \Lambda$, one should work in microscopic
%UV completed constructions and drop off the effective constructions.

%%%%%%%%%%%%%%%%%%%%%%%%%%%%%%%%%%%%%%%%%%%%%%%%%%%%%%%
%%%%%%%%%%%%%%%%%%%%%%%%%%%%%%%%%%%%%%%%%%%%%%%%%%%%%%%%

\noindent
We also want to underline the main difference with an infrared-dominated ``freeze in" scenario, where the DM is also absent 
in the early Universe. Indeed, in orthodox
freeze-in, the relic abundance increases very slowly as a function of $m_\chi/T$, and the process which populates
the Universe with DM is frozen at the time when the temperature drops below the mass of the dark
matter, Boltzmann-suppressing its production by the thermal bath, which does not have 
sufficient energy to create it through annihilation. This can be considered as a fine tuning: the relic abundance
should reach the WMAP value at a definite time, $T \simeq m_\chi/3$. In a sense, it is a common feature among
freeze--in and freeze--out scenarios. In both cases the fundamental energy scale which stops the (de)population process
is $m_\chi/T$. When the mediator mass $m_{Z'}$ is larger than the reheating temperature, the fundamental scale which determines
the relic abundance is $T_{\rm RH} / m_{Z'}$ or $T_{\rm{RH}}/\Lambda$ in the effective approach. 
 The DM abundance is then saturated from the beginning, at the reheating time, and thus stays constant during
 the rest of the thermal history of the Universe, and is nearly independent of the mass of the DM:
 no fine tuning is required, and no ``special" freeze-in at $T \simeq m_\chi/3$. This is a particular case of the 
 NETDM framework presented in \cite{Mambrini:2013iaa}. Furthermore, the NETDM mecanism has the interesting properties to avoid large thermal corrections to dark matter mass. The reason is that all dark sector particles are approximately decoupled from the visible medium of the Universe.\footnote{While the thermal masses of visible particles may change the DM production rate, we have checked that this effect is negligible.}

%=====================================   RESULTS MZ' < TRH   ================================================

\subsection{$m_{Z'} < T_{\mathrm{RH}}$}

\subsubsection{Generalities}

\noindent
The case of light mediators (in comparison to the reheating temperature) is more complex and rises several specific issues.
We concentrate in this section on the computation of the DM relic abundance in the kinetic-mixing 
framework because it can be easily embedded in several ultraviolet completions. However, our analysis is valid for any kind of 
models with an extra $U(1)$ gauge group. The kinetic mixing
$\delta$ is indeed completely equivalent to an extra $U(1)$  millicharge  for the visible sector and one can think $\delta$ as the  charge of 
the SM particles (visible world) to the $Z'$.
Cosmological constraints allow us to restrict the parameter space of the model in the plane ($\delta, m_{Z'}, m_{\chi})$.
However we should consider two options for the mediator $Z'$: either it is in thermal equilibrium with the SM plasma,
or, in analogy with the DM,  it has not been appreciably produced during the reheating phase. 

 %A priori one can make two assumptions for $Z'$: either we state that $Z'$ was present in the thermal bath at the end of reheating epoch, similar to the rest of SM particles, or, on the contrary, we state it was not, in the very same way we assume for the DM population (actually defining the non-equilibrium framework).  In fact, assuming that the $Z'$ particles were from the beginning in equilibrium with the thermal bath would imply a too large contribution to the DM relic abundance, as we estimate below.
 
 \noindent
 The differential equation for the decay process $Z'\to\bar\chi\chi$, in the case where the DM annihilation is neglected, can be expressed as:
\bea
&&
\frac{dY}{dx} = \frac{m_{Z'}^3 \Gamma_{Z'} g_{Z'}}{2 \pi^2H x^2 {\bf s}} K_1(x).
\eea
\noindent
where $x\equiv m_{Z'}/T$, $\Gamma_{Z'}$ the decay width of $Z'$ and $g_{Z'} =3$ giving the degree of freedom of the massive gauge boson $Z'$. Expressing the entropy and Hubble parameter as:
\be
{\bf s} = g_*^s \frac{2 \pi^2}{45} \frac{m_{Z'}^3}{x^3}, ~~~H= \sqrt{g_*}
\sqrt{\frac{4 \pi^3}{45}} \frac{m_{Z'}^2}{x^2 M_p}
\nonumber
\ee

\noindent
 we finally obtain the equation
\be
Y_0 \approx\left( \frac{45}{\pi} \right)^{3 \over 2}
\frac{1}{g_*^s \sqrt{g_*}} \frac{M_p \Gamma_{Z'} g_{Z'}}{8 \pi^4 m_{Z'}^2}
\int_{\frac{m_{Z'}}{T_{RH}}}^\infty
x^3 K_1(x) dx.
\ee

\noindent
Approximating $\Gamma_{Z'} \simeq q^2_Dg_D^2 m_{Z'}/(16 \pi)$, $q_D g_D$ being the
effective gauge coupling of $Z'$ and DM, and also taking
 $g_*^s \simeq g_*$ at the energies of interest, we can write

\be
Y_0 \simeq \left(\frac{45}{\pi} \right)^{3/2} \frac{q_D^2 g_D^2 M_p}{128
\pi^5 m_{Z'}} \int_{\frac{m_{Z'}}{T_{RH}}}^\infty x^3 K_1(x) dx.
\label{Ydecay}
\ee

\noindent
%Using $\int_0^\infty x^3 K_1(x) dx \simeq 4.7$ 
%we have
%\be
% \int_{X}^\infty x^3 K_1(x) dx \simeq \sqrt{2 \pi}
%\int_{\sqrt{\frac{M_A}{T_{RH}}}}^\infty t^6 e^{-t^2} dt \simeq
%\sqrt{\frac{\pi}{2}} e^{-\frac{M_A}{T_{RH}}}
% \left( \frac{M_A}{T_{RH}} \right)^{5/2}
%\ee
%\be
 %\int_{\frac{m_{Z'}}{T_{RH}}}^\infty x^3 K_1(x) dx \simeq
%\sqrt{\frac{\pi}{2}} ~e^{-\frac{m_{Z'}}{T_{RH}}}
 %\left( \frac{m_{Z'}}{T_{RH}} \right)^{5/2}
%\ee

%\noindent
%Applying then 
Using $\int_0^\infty x^3 K_1(x) dx \simeq 4.7$  and Eq.(\ref{Om0}) we obtain

%\be
% \Omega_0 h^2\simeq \left\{ \begin{array}{ll}
%2 \times 10^{22}g_D^2 \frac{m_\chi}{m_{Z'}} & m_{Z'} << T_{RH}\\
%5 \times10^{21} g_D^2 \frac{m_\chi}{m_{Z'}}
%e^{-\frac{m_{Z'}}{T_{RH}}} \left( \frac{m_{Z'}}{T_{RH}} \right)^{3/2} & m_{Z'} >> T_{RH}\\
%\end{array} \right.
%\label{Om0dec}
%\ee

\be
\Omega_0 h^2\simeq 2 \times 10^{22}q_D^2g_D^2 \frac{m_\chi}{m_{Z'}}.
\label{Om0dec}
\ee

\noindent
To respect WMAP/Planck data in a FIMP scenario one thus needs $g_D\simeq 10^{-11} $ if $Z'$ is at TeV scale.
For much higher values of $g_D$, the DM joins the thermal equilibrium at a temperature $T \gg m_\chi$ 
and then recovers the classical freeze out scenario.

\noindent
Thus, a first important conclusion is that a $Z'$ in thermal equilibrium with the plasma and decaying dominantly to DM would naturally overpopulate the DM which would thus thermalise with plasma, ending up with the standard freeze-out history.
%a $Z'$ decaying dominantly to DM in thermal equilibrium with the plasma would naturally overpopulate the DM which would then enter in equilibrium with the SM plasma.} 
We then have no choice than to concentrate
on the alternative scenario where $Z'$, same as the DM, was not present after inflation. Thus the interaction 
of the SM bath (and the DM generated from it)
  could create it in a considerable amount. This is discussed below.
 
% So if we want to make sure that a thermal $Z'$  never contributes to DM population, we need to study the rate $\Gamma$ of the production process $\overline{\rm SM}{\rm SM}\to Z'$ and compare it to the expansion rate of the Universe, in order to check that, indeed, it never thermalises, thus being safe to further forget about $Z'\to\overline{\rm DM}{\rm DM}$. We present next this calculation.
 
\subsubsection{Chemical equilibrium of the dark sector}

If $Z'$ is generated largely enough at some point during the DM genesis, it will surely affect the DM final relic abundance through the efficient DM-$Z'$ interactions.
 In the study of the evolution of the $Z'$ population it may happen that $Z'$ enters in a state of chemical equilibrium exclusively with DM, independently of the 
 thermal SM bath, and thus with a different temperature. This ``dark thermalisation" can have some effect on the final DM number density. The analysis we perform 
 here is inspired from \cite{Chu:2011be}, which was however applied to a different model.

%The idea is to compute a temperature $T'=T'(T)$ which can be interpreted as a measure of the average kinetic energy $\langle K\rangle$ of DM and $Z'$. When chemical equilibrium between DM and $Z'$ is attained, $T'=T'_{\rm eq}$, where usually $T'_{\rm eq} < T$. In other words, the dark sector will normally reach an independent chemical equilibrium, at a temperature smaller than the temperature $T$ of the thermal bath. This chemical equilibrium between DM and $Z'$ could in principle affect the final $n_{\chi}$. Thus it is important to see its contribution. 

%To study the dark thermalisation the key assumption is to impose that DM and $Z'$ are always in kinetic equilibrium. This assumption is good when the number densities of DM and/or $Z'$ are abundant, because it means that normally DM particles will find $Z'$ partners to elastically scatter off. On the other hand it is not a good assumption when  $n_{Z'}$ is negligible. However, in this case the dark-to-dark processes are not going to affect appreciably the evolution of $n_\chi$. Summarising, the assumption is almost always good, and the regimes for which it is not good are not relevant to the $n_\chi$ evolution. 

If the $Z'-$DM scattering rate is larger than the Hubble expansion rate of the Universe\footnote{For a deeper analysis on this, see \cite{bfmz}.}, these two species naturally reach  kinetic equilibrium, with a well defined temperature $T'$, which a priori is different from (and is a function of) the thermal bath (photon) temperature, $T$. This temperature
  $T'$ increases slowly (given the feeble couplings) due to the transfer of energy from the thermal bath, which determines the energy 
  density $\rho'$ and pressure $P'$ of the dark sector. The Boltzmann equation governing the energy transfer in this case is:

\bea
&&\df{d\rho'}{dt}+3H(\rho'+P')=\int\prod_{i=1}^4 d^3\bar p_i 
f_1(p_1)f_2(p_2) \nonumber \\
&&\times |{\cal M}|^2 (2\pi)^4\delta^{(4)}(p_{\rm in}-p_{\rm out})\cdot E_{\text{trans.}}\nonumber\\
&&= \df{1}{2048\pi^6}\int^\infty_{4m_{\chi}^2}ds K_2(\df{\sqrt{s}}{T})T\sqrt{(s-4m_{\chi}^2)s}
|\tilde {\cal M}_{12\rightarrow \chi\bar\chi}|^2\nonumber\\
&&+ \df{1}{128\pi^4}K_2(\df{m_{Z'}}{T})m_{Z'}T\sqrt{m_{Z'}^2-4m_{1}^2}|\tilde{\cal M}_{Z'\to12}|^2~,
\label{rhop1}
\eea
\newline
where 1 and 2 are the initial SM particles and $m_1=m_2$, $|\tilde {\cal M}|^2$s
have been defined below Eq.(\ref{Boltz1}) summing over all initial and final degrees of freedom. For SM pair annihilation, the energy transfer per
collision is $E_{\text{trans.}}= E_1+E_2$. It can be useful to write an analytical approximation for the solution $\rho'(T)$ in the early Universe.
Indeed  for $T \gg m_{Z',\chi}$, it is easy to show that Eq.(\ref{rhop1}) reduces to
 
 \bea
 &&
 \frac{d (\rho' / \rho)}{d T} \simeq - 640 \sqrt{\frac{45}{\pi}} \frac{\alpha \delta^2 M_p}{\pi^7 T^2 g_*^{3/2}} 
 \nonumber
 \\
 &&
 \Rightarrow ~
  \left( \frac{T'}{1~\mathrm{GeV}}  \right) \simeq 3000 ~\sqrt{\delta} \left( \frac{T}{1 ~\mathrm{GeV}} \right)^{3/4}
  \label{Eq:rhop2}
 \eea

\noindent
supposing that the dark bath is in kinetic equilibrium ($\rho' \propto (T')^4$) with $\alpha=g^2/4 \pi$
 (see next section for more details). 
 Even if all our analysis was made using the analytical solutions of the coupled Boltzmann system,
 we checked that this analytical solution is a quite good approximation to the exact numerical
 solution of Eq.(\ref{rhop1}) and will be very useful to understand the physical phenomena hidden by the numerical results.

\noindent
While presenting a detailed study of the visible-to-dark energy transfer is out of the scope of this work, we just want to point out that there is
 typically a moment at which the dark sector (i.e. DM plus $Z'$) is sufficiently populated as for creating particles out of itself, e.g. in
  processes as a t-channelled $\chi\bar\chi\to Z'Z'\to 2\chi 2\bar\chi$. As this happens out of a total available energy $\rho'$ at any given time,
   the net effect is to increase $n_\chi$ and $n_{Z'}$ at the cost of decreasing $T'$. 

\begin{figure}[ht]
\includegraphics[width=0.5\textwidth]{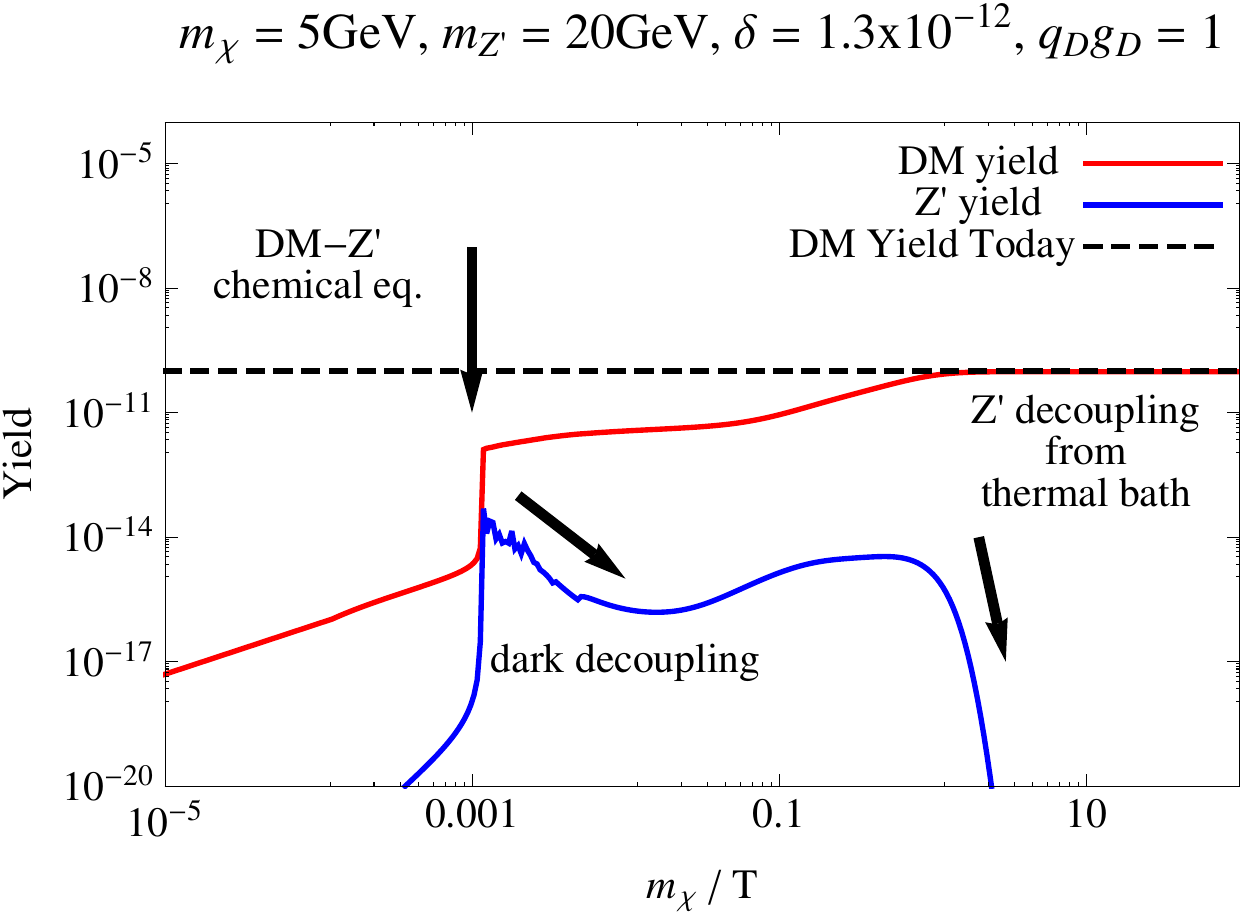}
\caption{\footnotesize{ Evolution of the yield for DM (red) and $Z'$ (blue) as a function of temperature for $m_{Z'} >2 m_\chi$.
{The set of parameters is given on the figure}.}} 
\label{fig:YvsX}
\end{figure}

\begin{figure}[ht]
\includegraphics[width=0.5\textwidth]{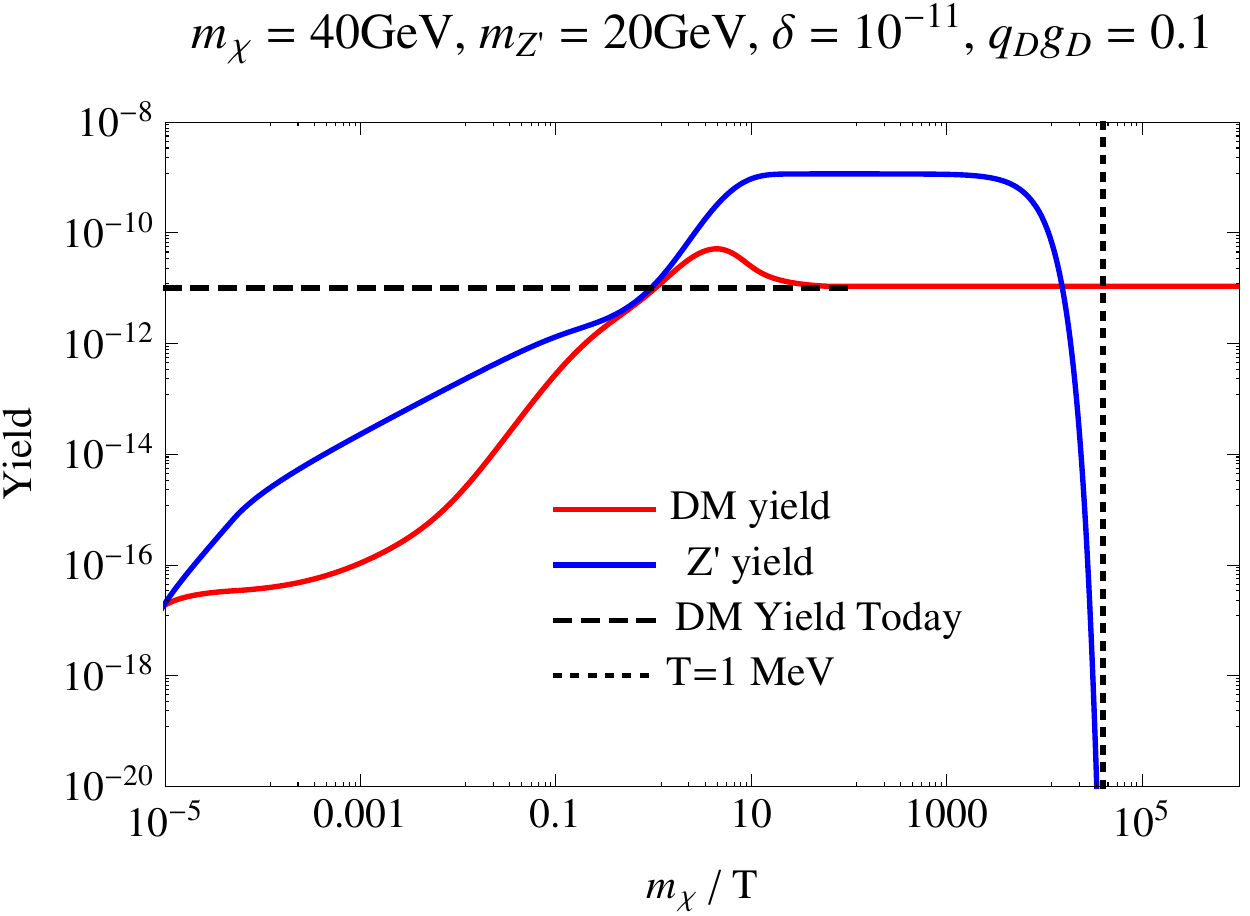}
\caption{\footnotesize{ Same as Fig.~(\ref{fig:YvsX}) with $m_{Z'} < 2m_\chi$. Note here a smaller $q_Dg_D$ is adopted to avoid too many dark matter annihilations.  }} 
\label{fig:YvsXbis}
\end{figure}

\noindent
To quantify the effect of DM-$Z'$ chemical equilibrium on the number densities of both particles, we solved the coupled set of their
 respective Boltzmann equations  (see appendix \ref{sec:Boltzmann}). The relevant $Z'$ production process is the scattering 
 $\overline{\chi} \chi \to Z' Z'$ (as compared to $\overline{\chi} \chi \to Z'$), whereas the relevant $Z'$ depletion process is the 
 decay $Z'\to \overline{\chi}\chi$ (as compared to $Z' Z'\to \overline{\chi} \chi$), but of course we have considered all the
  processes when solving the Boltzmann equations. 
The results are shown in Fig.~(\ref{fig:YvsX}) for $m_{Z'} > 2 m_\chi$ and in Fig.~(\ref{fig:YvsXbis}) for $m_{Z'} < 2 m_\chi$.

%%%%%%%%%%%%%%%%%%%%%%%%%%%%%%%%%%%%%%%%%%%%%%%%%%%%%%%%
%%%%%%%%%%%%%%%%%%%%%%%%%%%%%%%%%%%%%%

%%%%%%%%%%%%%%%%%%%%%%%%%%%%%%%%%%%%%%%%%%%%%%%%%%%%%%%%
%%%%%%%%%%%%%%%%%%%%%%%%%%%%%%%%%%%%%%

\noindent
Figure \ref{fig:YvsX} presents several original and  interesting features. We can separate the thermal events in 4 phases that we detail below:
dark kinetic equilibrium of the dark matter candidate, self exponential production of dark matter through its annihilation, decoupling
of the $Z'$ from the dark bath and then decoupling of $\chi$ and $Z'$ from the thermal standard bath.

\noindent
Indeed, we can notice a first kind of plateau for the dark matter yield 
$Y_{\chi}$ at $T \gg 10^{3}$ GeV. This corresponds to the time when the dark matter concentration is sufficient  to enter    equilibrium
with itself through the exchange of a virtual $Z'$ ($s$ or $t$ channel). Indeed, the condition $n_\chi \langle \sigma v \rangle > H(T)$ can be expressed as

\bea
&&\left\{10^{-5} M_p g_*^s \delta^2 \alpha~ T^2 \right\}
\times
\df{(q_D g_D)^4}{(4\pi)^2 T^2} > \df{1.66}{M_p}\sqrt{g_*}T^2 \nonumber\\
&&\Rightarrow
T \lesssim 1.6\times 10^{15} g_*^{1/4}\alpha^{1/2} \delta ~~{\rm GeV} 
\eea

\noindent
where we have used an approximate solution of Eq.(\ref{Boltz1}) at high temperatures:

\be
Y_\chi \simeq \alpha ~ \delta^2 ~ \frac{10^{14} ~\mathrm{GeV}}{T}
\ee

\noindent
with $\alpha = \frac{g^2}{4 \pi}$. The result is then in accordance with what we observed numerically.

\noindent
We then observe in a second phase, around $m_\chi/T=10^{-3}$, a simultaneous and sharp rise in the number density of DM and $Z'$. This 
is because the dark sector enters in a phase of chemical equilibrium with itself, causing the population of both species to increase.
Moreover, in the case $m_{Z'}> 2 m_\chi$, we observe that the width of the $Z'$ $\Gamma_{Z'}$ is much larger than the production
rate through the $t$ channel $\chi \chi \rightarrow Z' Z'$:
\bea
\Gamma_{Z'} &\simeq& \frac{(q_Dg_D)^2 }{16\pi} m_{Z'} \simeq 0.4 ~\mathrm{GeV}~, 
\\
n \langle \sigma v \rangle_{\chi \chi \rightarrow Z' Z'} &\simeq& 10^{12} g_*^s ~  \delta^2  \alpha ~(q_D g_D)^4 
 \nonumber \\
 &\simeq&10^{-12} \sqrt{\df{T}{1{\rm GeV}}} ~{\rm GeV}.
\nonumber
\eea 

\noindent
In other words, as soon as a $Z'$ is produced, it automatically decays into two DM particles before having the time
to thermalise or annihilate again. We then observe an exponential production of DM. Of course, each
product of the $Z'$ decay possesses half of the initial energy of the annihilating DM, this energy also decreasing
exponentially. As a consequence, 
 the temperature of the dark sector, $T'$, typically
 drops below $m_{Z'}$ at a certain temperature $T$ such that the dark sector does not have enough energy for maintaining 
 an efficient $Z'$ production\footnote{Strictly speaking
 one should not use the word $temperature$ $T'$ during this very short time but more express ourselves in terms of energy.}. 
 This is illustrated as ``dark decoupling" in Fig.~(\ref{fig:YvsX}), where the excess of $Z'$ population decays mostly to DM particles. 
 We can understand this phenomenon by looking more in details at the  solution of the transfer of energy (\ref{Eq:rhop2}). 
 Taking $T' \simeq m_{Z'}$ in Eq.(\ref{Eq:rhop2}), we can check that the decoupling of the $Z'$ from the dark bath happens around
  a temperature $T \simeq 2$ TeV when the DM does not possess sufficiently energy to produce
  a $Z'$ pair. This result is in accordance with the value observed in Fig.~(\ref{fig:YvsX}) along the arrow
  labelled {\it dark decoupling}.

\noindent 
 However, the thermal (standard) bath is still able to slowly produce $Z'$ after its decoupling from the dark bath but at a very slow
 rate (proportional to $\delta^2$) up to the moment at which the temperature $T$ drops below $m_{Z'}$, 
 when the $Z'$ population decays completely as we can also observe in Fig.~(\ref{fig:YvsX}). During this time the DM population
 increases also slowly due to the annihilation of SM particles through the exchange of a virtual $Z'$  added to the product of the $Z'$ decay
  until $T$ reaches $m_{\chi}$.
  
  \noindent
  We also depict in Fig.~(\ref{fig:YvsXbis}) the evolution of the $Z'$ and DM yields in the case $m_{Z'} < 2~m_{\chi}$.
  We observe similar features, except that the $Z'$ does not decouple from the dark bath and is not responsible anymore for the exponential
  production of DM. The DM decouples first from
  the plasma, and then the $Z'$ continues to be produced at a slow rate, being also largely populated by the $t-$channel annihilation
  of the dark matter. However, it never reaches the thermal equilibrium with the thermal
  bath as it decays to SM particles (at a very low rate proportional to $\delta^2$) at a temperature of about 1 MeV, not affecting the
  primordial nucleosynthesis (see below for details).
 
 % Concerning the DM population, it received a boost due to the chemical equilibrium
 % with $Z'$ during that short period, causing $Y_{\chi}$ to arrive sooner to its value today, with respect to the case where no $Z'$-DM 
  %thermalisation is taken into account.  

%After this moment both DM (fig. \ref{fig:YvsX}-top) and $Z'$ (fig. \ref{fig:YvsX}-bottom) evolve exclusively according their separate interactions with the thermal bath, similar to the evolution before the stage of chemical equilibrium. 

\subsubsection{Cosmological constraints}
\label{sec:Constraints}

\begin{figure}[ht]
\centering
\includegraphics[width=0.5\textwidth,angle=0]{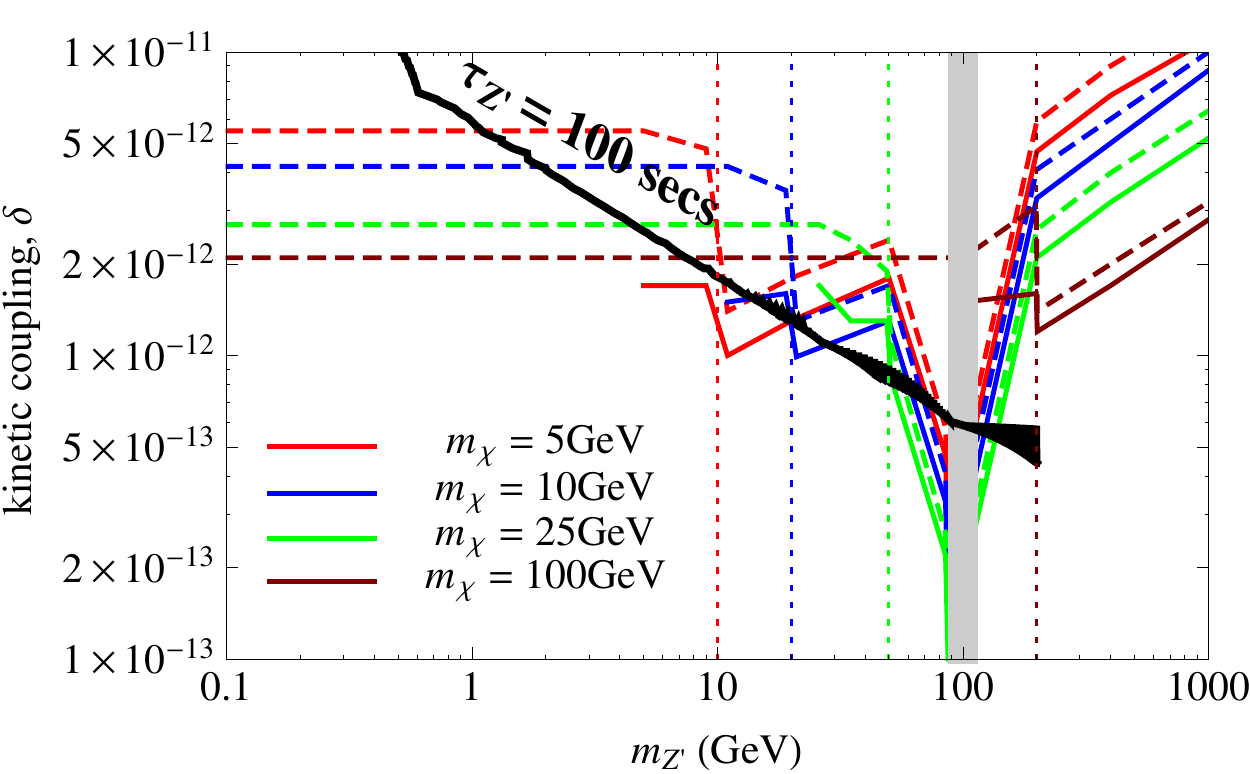}
\caption{\footnotesize{ Kinetic-mixing coupling $\delta$ as a function of $m_{Z'}$ for different values of $m_\chi$: 5, 10, 25 and 100 GeV for red,
 blue, green and brown curves, respectively. These lines are in agreement with WMAP: $\Omega_\chi h^2 \sim 0.12$. We have fixed 
 $q_D g_D = 1$, as before. Solid lines are obtained taking into account the ``dark thermalisation" effect (see text for details) whereas 
 dashed lines are obtained without such an effect.  The solid black line shows BBN constraints (see text details),
which apply, for each DM mass (shown with dotted lines), to the region $m_{Z'} < 2m_\chi$.}}
\label{fig:km1}
\end{figure}

 \noindent
 The PLANCK collaboration \cite{Ade:2013zuv} recently released its results and confirmed the WMAP \cite{WMAP} non--baryonic content of the Universe.
 It is then important to study in the ($m_\chi, m_{Z'}, \delta$) parameter space the region which is still allowed by the cosmological WMAP/PLANCK constraint.
 As we discussed in the previous section, a small kinetic mixing can be sufficient to generate sufficient relic abundance.
  We show in Fig.~(\ref{fig:km1}) the plane ($\delta$,$m_{Z'}$) compatible with WMAP/PLANCK data ($\Omega h^2 \simeq 0.12$)
   for different dark matter masses. Depending on the relative value between $m_\chi$ and $m_{Z'}$, we can distinguish
   four regimes clearly visible in Fig.~(\ref{fig:km1}):

%For the results shown below, we will assume that $Z'$ was not, from the beginning, part of the thermal bath, so its decay does not contribute to the DM genesis (as we checked in last section). 

%%%%%%%%%%%%%%%%%%%%%%%%%%%%%%%%%%%%%%%%%%%%%%%%%%%%%%%%
%%%%%%%%%%%%%%%%%%%%%%%%%%%%%%%%%%%%%%

\begin{enumerate}[(a)]
\item $m_{Z'}<2m_\chi$. 
In this regime, the dark matter is mainly produced from the plasma through $s-$channel exchange of the $Z'$ and then decouples
from the thermal bath at $T \simeq m_\chi$. Dark matter then annihilates into two $Z'$ through $t-$channel process if kinetically allowed (see Fig.~\ref{fig:YvsXbis}). 
For light $Z'$, the amplitude of dark matter production\footnote{In this region the $Z'$-SM couplings (see Appendix A) 
are roughly proportional to $\delta$, since $\sin\phi\ll\delta$ for 
 the values of $\delta$ and $m_{Z'}$ in consideration.} 
 ($|{\cal M}|^2 \propto \delta^2 m_\chi^2/s \sim \delta^2 m_\chi^2/T^2$ from Eq.\ref{procKM3}) 
 and the annihilating rate ($\chi \chi \rightarrow Z'Z'$) after the decoupling time
are both independent of $m_{Z'}$. As a consequence, the relic abundance is also independent of $m_{Z'}$ (but strongly dependent of $\delta$)
 as one can observe in the left region of Fig.(\ref{fig:km1}).

\item $2m_{\chi}< m_{Z'} < M_Z$. 
We notice a sharp decrease in the  values of $\delta$ occurring around $m_{Z'}= 2 m_\chi$. Indeed, for $m_{Z'} > 2 m_\chi$
there exists a temperature in the plasma for which the resonant production of onshell $Z'$ is abundant ($T \simeq m_{Z'}/2$).
The $Z'$ being unstable, it immediately decays into 2 dark matter particles increasing its abundance.  
The rate of the dark matter production from the standard model bath around the pole $T\simeq m_{Z'}/2$
is proportional to $\delta^2 m^2_\chi T^2/ m^2_{Z'} \Gamma^2_{Z'}$ (Eq.\ref{procKM3}). This rate is higher
than in the region $m_{Z'} < 2 m_\chi$ where $|{\cal M}|^2 \propto  \delta^2 m_\chi^2/T^2$: $\delta$
should then be smaller in order to still respect PLANCK/WMAP constraint.

%The resonance may be accessible in some cases, and the DM yield is essentially proportional
 %to$\delta^2/\Gamma_{Z'}$. The on-shell $Z'$ contribution tends to increase the 
% amplitude, so $\delta$ is forced to decrease in the right amount as to maintain the relic density's correct value. 

\item $m_{Z'}\approx M_Z$. This is the region of maximal mixing: $\phi\approx \pi/4$. The total amplitude of annihilation in
 Eq.(\ref{procKM1}) is maximised,  driving $\delta$ toward very small values in order to respect PLANCK/WMAP constraint. 
However, this region is excluded by electroweak measurements because of large excess in the $\rho$ parameter 
(see \cite{Mambrini:2010yp} for a complete analysis in this regime).

\item $2m_{\chi}< M_{Z} < m_{Z'}$. For even larger values of $m_{Z'}$ the amplitude has a smooth tendency of decreasing with 
$m_{Z'}$ from its dependence on the width. The majority of the dark matter population is indeed created when the temperature
of the universe, playing the role of a statistical accelerator with time dependent centre of mass energy, reaches
$T \simeq m_{Z'}/2$ (or $m_{Z}/2$). The production cross section through $s-$channel exchange of $Z'$ is then proportional
 to $\delta^2/m^2_{Z'}\Gamma^2_{Z'} \propto \delta^2/m^4_{Z'}$. Keeping constant final relic abundance implies 
 $\delta^2/m^4_{Z'}=$ constant, which is observed in the right region of Fig.(\ref{fig:km1}).
\end{enumerate}

\noindent
For the sake of completeness, we also show in Fig.~(\ref{fig:km1}) the effect of allowing the $Z'$ and dark matter to enter in a phase of 
chemical equilibrium (solid lines), see Fig.~(\ref{fig:YvsX}) and compare it  to the more naive case where no dark-thermalisation is
 taken into account (dashed lines).  We observe that depending on the DM and $Z'$ masses, the correction caused by the 
 dark-thermalisation for $q_Dg_D=1$ is at most a factor 2.

 \noindent
 Meanwhile, a general look at Fig.~(\ref{fig:km1}) tells us that the order of magnitude of $\delta$ to respect relic abundance data is
  generally in the range $10^{-12}$--$10^{-11}$, which is in absolute value of the same order that typical FIMP couplings
  obtained in the literature for different frameworks \cite{Hall:2009bx,Chu:2011be,McDonald:2001vt,Yaguna:2011qn,Arcadi:2013aba}
  but with a much richer phenomenology due to the instability of the mediator and the existence of dark thermalisation.  
  It is interesting to note that such tiny kinetic mixing, exponentially suppressed, 
  is predicted by recent work on higher dimensional compactification and string phenomenology to lie within
   the range $10^{-12}\lesssim \delta \lesssim 10^{-10}$ \cite{Javier}.
  
\noindent
Finally, due to the feeble coupling $\delta$, it is important  to check constraints coming from Big Bang
Nucleosynthesis (BBN) in the specific case $m_{Z'} < 2 m_\chi$. Indeed, if $Z'$ is lighter than the dark matter,
 the $Z'$ will slowly decay to the particles of the thermal bath, potentially
affecting the abundance of light elements. For the ranges of $Z'$ masses
we consider here, a naive bound from BBN can be obtained by simply
requiring the $Z'$ lifetime to be shorter than ${\cal O}(100)$ seconds. This is translated
into a lower bound on the kinetic coupling $\delta$, represented by the
black solid line in Fig.~(\ref{fig:km1}), where the bound applies, for every $m_\chi$
(see dotted lines), to the region $m_{Z'}<2m_\chi$. We see how the BBN
bounds strongly constrain the region of lightest $Z'$, $m_{Z'}\lesssim 1$ GeV for the DM masses considered here.
A more detailed study of nucleosynthesis processes in this framework can be interesting but is far beyond the scope of this paper

%=====================================================================================================================

\subsubsection{Other constraints}

In \cite{Mambrini:2010yp}  several low-energy processes have been used in order to constrain the parameter space of the model we analysed.  
We refer the reader to that work in order to see the study in more  details. In this section, we just want to extract one of the strongest bounds, which 
comes from Electroweak Precision Tests (EWPT). Indeed, since the model modifies the coupling of the $Z$ to all fermions, 
the decay rate to leptons, for example, is in principle modified.  It turns out that a model is compatible with EWPT under the condition

\be
\label{EWPT}
\left(\df{\delta}{0.1}\right)^2\left(\df{250{\rm GeV}}{m_{Z'}}\right)^2 \lesssim 1~.
\ee 

\noindent
For a very light $Z'$ of $m_{Z'}\sim 1$ GeV, the EWPT constraints require $\delta\lesssim 10^{-4}$, which is well above the 
WMAP constraints shown in Fig.~(\ref{fig:km1}).  Also, since the model modifies the $Z$ mass, constraints coming from the 
deviation of the SM prediction  for the parameter $\rho\equiv M_W^2/M^2_Zc^2_W$ are also expected to appear; 
however, they turn out to be weaker or similar to those of EWPT.

\noindent
Direct Detection experiments, leaded by XENON \cite{Aprile:2011hi}, are able to put much more stronger bounds on the model. The dark matter 
candidate can scatter off a nucleus 
through a $t$-channel exchange of $Z$ or $Z'$ bosons (see e.g. \cite{Mambrini:2010yp}\cite{Arcadi:2013qia}). It turns out that for the dark matter and
 $Z'$ masses considered, the  XENON1T analysis is expected to push $\delta$ to values $\delta \lesssim 10^{-4}$, to say the strongest. Again
  here those bounds are not competitive with those shown in Fig.~(\ref{fig:km1}).

\noindent
As an example of constraints coming from indirect detection, we can use synchrotron data. The dark matter particles in the region of the 
Galactic Centre can annihilate
 to produce electrons and positrons, which will emit synchrotron radiation as they propagate through the magnetic fields of the galaxy. In \cite{Mambrini:2012ue}
 the authors constrain the kinetic mixing in the framework of freeze-out.
 The synchrotron data is able to put bounds on the parameter space of the model, provided that $m_\chi$ and $m_{Z'}$ are light enough 
  (less than ${\cal O}$(100) GeV), and for  values of $\delta$ compatible with a thermal relic which are much larger 
  than those required to fit a WMAP with a froze-in dark matter. So given the small $\delta$ values considered here, the synchrotron bounds are unconstraining.

\section{Conclusions}

\noindent
In this work we have studied the genesis of dark matter by a $Z'$ portal for a spectrum of $Z'$ mass from above the reheating temperature down to a
few GeV. Specifically, we have distinguished two regimes: 1) a very massive portal whose mass is above the reheating 
temperature $T_{RH}$, illustrated by effective, vector-like interactions 
between the SM fermions and the DM, and 2) a weak-like portal, illustrated by a kinetic-mixing model with an extra $U(1)$ boson, $Z'$, which 
couples feebly to the SM but with unsuppressed couplings to the dark matter, similar to a secluded dark sector.

\noindent
In the case of very massive portal we solved the system of Boltzmann equations and obtained the expected dependance of the 
dark matter production with the reheating temperature. By requiring consistency with the WMAP/PLANCK's measurements of the non--baryonic 
relic abundance, the scale of the effective interaction $\Lambda$ should be approximatively $\Lambda \simeq10^{12}$ GeV, for $T_{RH}\approx 10^9$ GeV.  

\noindent
For lighter $Z'$ that couples to the standard model through its kinetic mixing with the standard model $U(1)$ gauge field, 
we considered $Z'$ masses in the 1 GeV--1 TeV range. The values of the kinetic mixing 
$\delta$ compatible with the relic abundance we obtained are $10^{-12}\lesssim \delta \lesssim10^{-11}$ 
depending on the value of the $Z'$ mass. For such values, the constraints coming from other experimental fields  
like direct or indirect detection and LHC production, become meaningless.
 However the bounds coming from the Big Bang nucleosynthesis can be quite important. 
  For the study of the dark matter number density evolution, we looked at the effect of  
  chemical equilibrium between dark matter and $Z'$ on the final
   dark matter population, which turns out for the parameter space we considered to give a correction of at most a factor of 2.

\section*{Acknowledgements}
 The authors would like  to thank
E. Dudas, A. Falkowski, K. Olive, M. Goodsell, T. Hambye, M. Tytgat, E. Fernandez-Martinez and M. Blennow for very useful discussions.
This  work was supported by
the French ANR TAPDMS {\bf ANR-09-JCJC-0146}  and the Spanish MICINNs
Consolider-Ingenio 2010 Programme  under grant  Multi- Dark {\bf CSD2009-00064}. X.C. acknowledges the  support of the FNRS-FRS, the IISN, the Belgian Science Policy (IAP VI-11).
Y.M.  and J.Q. acknowledge partial support from the European Union FP7 ITN INVISIBLES (Marie
Curie Actions, PITN- GA-2011- 289442), the ERC advanced grant Higgs@LHC and thanks the Galileo Galilei Institute for Theoretical Physics
for the hospitality and the INFN for partial support during the completion of this work.
B.Z. acknowledges the support of MICINN, Spain, under the contract FPA2010-17747, as well as the hospitality of LPT, Orsay during the completion of this project.

\newpage
\appendix
%\section{Bla} 
\section{Boltzmann equations}
\label{sec:Boltzmann}

The relevant processes happening between the dark sector and SM\footnote{Here
  we are not writting the contributions from processes like
  $SM\gamma \rightarrow SM Z'$ and $SM\overline{SM} \rightarrow \gamma
  Z'$; but they are taken into account for the numerical analysis.}, and with itself, are:\\
\begin{itemize}
\item $a$: $SM \overline{SM}\to Z'$, and $\bar a$:  $SM \overline{SM}\leftarrow Z'$
\item $b$: $\chi \overline{\chi}\to Z'$, and $\bar b$:  $\chi \overline{\chi}\leftarrow Z'$
\item $c$: $Z' Z' \to \chi \overline{\chi}$, and $\bar c$: $Z' Z' \leftarrow \chi \overline{\chi}$
\item $d$: $\chi \overline{\chi}\to SM \overline{SM}$, and \\$\bar d$: $\chi \overline{\chi}\leftarrow SM \overline{SM}$.
\end{itemize}
The Boltzmann equations for the $Z'$ and DM comoving number densities are:
\bea
\df{dY_{Z'}}{dT} &=& \df{1}{HT}\left[ \Gamma_{\bar a} (Y^{eq}_{Z'} - Y_{Z'}) - \Gamma_{\bar b} Y_{Z'} \right. \\
&+& \sv_{b} Y_{\chi}^2 {\bf s}
- \left. \sv_{c}Y_{Z'}^2{\bf s} + 2\sv_{\bar c}Y_\chi^2{\bf s}  \right] \nonumber
 \label{YZ}
\eea
\bea
\label{Ychi}
\df{dY_\chi}{dT} &=& \df{1}{HT}\left[ \sv_{d}((Y_{\chi}^{eq})^2- Y_\chi^2){\bf s} - \sv_{b}Y^2_{\chi}{\bf s} \right.\\
&+& 
\Gamma_{\bar b} Y_{Z'} 
-\left. 2\sv_{\bar c}Y^2_{\chi}{\bf s} + \sv_{c}Y^2_{Z'}{\bf s} \right]. \nonumber 
\eea
Here in Eq.(\ref{YZ}), in the very first term, we have made use of the chemical equilibrium condition for a process $A\leftrightarrow B \bar B$
\[
\sv_{BB\to A} (Y^{eq}_{B})^2~{\bf s} = \Gamma_{A\to BB} Y^{eq}_A~.
\]

Besides, in Eq.(\ref{Ychi}), the term proportional to $\sv_{d}$ does not
contain the contribution from on-shell $Z'$, because it is already included in
the term going with $\Gamma_{\bar b}$. The reason for this, is that the
typical time the reaction $SM\overline{SM}\leftrightarrow \chi\overline{\chi}$
takes to happen, is $t_{\rm typ}$. This period, even if usually very short, is large enough as to consider $t_{\rm typ}\gtrsim dt$, where $dt$ is the characteristic time interval when solving the Boltzmann equation. In other words, the evolution dictated by the Boltzmann equation is such that there are always physical (on-shell) $Z'$ particles around, which effectively contribute to a $Z'$ decay.

The Boltzmann equation describing the evolution of the energy density transferred from the SM to the dark sector is:

\bea
&&\df{d\rho'}{dt}+3H(\rho'+P')=\int\prod_{i=1}^4 d^3\bar p_i
f_1(p_1)f_2(p_2) \nonumber \\
&&\times |{\cal M}|^2 (2\pi)^4\delta^{(4)}(p_{\rm in}-p_{\rm out})\cdot E_{\text{trans.}}\nonumber\\
&&= \df{1}{2048\pi^6}\int^\infty_{4m_{\chi}^2}ds K_2(\df{\sqrt{s}}{T})T\sqrt{(s-4m_{\chi}^2)s}
|\tilde {\cal M}_{12\rightarrow \chi\bar\chi}|^2\nonumber\\
&&+ \df{1}{128\pi^4}K_2(\df{m_{Z'}}{T})m_{Z'}T\sqrt{m_{Z'}^2-4m_{1}^2}|\tilde{\cal M}_{Z'\to12}|^2~,
\label{rhop}
\eea
\newline
where 1 and 2 are the initial SM particles and $m_1=m_2$, $|\tilde {\cal M}|^2$s
have been defined below Eq.(\ref{Boltz1}) summing over all initial and final degrees of freedom. For SM pair annihilation, the energy transfer per
collision $E_{\text{trans.}}= E_1+E_2$. The pressure $P'$ is: 
\bea
&& P'=\rho'_{\rm rel}/3 ~,\nonumber \\
&&\rho'_{\rm rel} = \rho' - 2n_{\chi}m_{\chi}-n_{Z'}m_{Z'} ~,
\eea
where $\rho'_{\rm rel}$ is the relativistic contribution to the energy density $\rho'$.

\section{Couplings in kinetic mixing model}
\label{app:coupl}
In this appendix we show the couplings of fermions (including DM) to the $Z$ and $Z'$ bosons in our model.
\newline\newline
The left ($L$) and right ($R$) couplings to the $Z$ boson are:
 \bea
&&(c_L)_f = - \df{(2g^2 T_{f_L}  - g'^2 Y_{f_L} )}{2\sqrt{g'^2+g^2}}\cos\phi  - \df{g'}{2} Y_{f_L}\sin\phi~\delta~,\nonumber \\
&& (c_R)_f =  \df{1}{2} g' Y_{f_R} \left(\df{g'}{\sqrt{g'^2+g^2}}\cos\phi -\sin\phi~\delta\right)~,
\eea
for SM fermions $f$, and 
\be
c_\chi = q_D g_D \sin\phi
\ee
for the DM. Similarly, the couplings to the $Z'$ boson to the SM fermions and DM $\chi$ are:
 \bea
&&(c_L)'_f = - \df{(2g^2 T_{f_L}  - g'^2 Y_{f_L} )}{2\sqrt{g'^2+g^2}}\sin\phi  +	 \df{g'}{2} Y_{f_L}\cos\phi~\delta~,\nonumber \\
&& (c_R)'_f =  \df{1}{2} g' Y_{f_R} \left(\df{g'}{\sqrt{g'^2+g^2}}\sin\phi +\cos\phi~\delta\right)~,\nonumber\\
&& c'_\chi =  q_D g_D \cos\phi~.
\eea
\newline
%).


\begin{thebibliography}{99}











%%%%%%%%%%%%%%%%%%%%%%%%%%%%%%%%%%%%%%%%%%%%%%%%






%====================== INTRODUCTION  ================================


\bibitem{Ade:2013zuv}
  P.~A.~R.~Ade {\it et al.}  [Planck Collaboration],
  %``Planck 2013 results. XVI. Cosmological parameters,''
  arXiv:1303.5076 [astro-ph.CO].
  %%CITATION = ARXIV:1303.5076;%%
  %155 citations counted in INSPIRE as of 22 May 2013
  
 
%\cite{Bergstrom:2000pn}
\bibitem{Bergstrom:2000pn}
  L.~Bergstrom,
%  ``Nonbaryonic dark matter: Observational evidence and detection methods,''
  Rept.\ Prog.\ Phys.\  {\bf 63} (2000) 793
  [hep-ph/0002126].
  %%CITATION = HEP-PH/0002126;%%
  %410 citations counted in INSPIRE as of 17 Feb 2013
  
 %\cite{Bertone:2004pz}
\bibitem{Bertone:2004pz}
  G.~Bertone, D.~Hooper and J.~Silk,
%  ``Particle dark matter: Evidence, candidates and constraints,''
  Phys.\ Rept.\  {\bf 405} (2005) 279
  [hep-ph/0404175].
  %%CITATION = HEP-PH/0404175;%%
  %1397 citations counted in INSPIRE as of 17 Feb 2013
  
  
   \bibitem{WMAP}
   G.~Hinshaw, D.~Larson, E.~Komatsu, D.~N.~Spergel, C.~L.~Bennett, J.~Dunkley, M.~R.~Nolta and M.~Halpern {\it et al.},
  %``Nine-Year Wilkinson Microwave Anisotropy Probe (WMAP) Observations: Cosmological Parameter Results,''
  arXiv:1212.5226 [astro-ph.CO].
  %%CITATION = ARXIV:1212.5226;%%
  
  
  

  
  
  %================================  LHC   ========================================
  
  
  \bibitem{Dreiner:2013vla}
  H.~Dreiner, D.~Schmeier and J.~Tattersall,
  %``Contact Interactions Probe Effective Dark Matter Models at the LHC,''
  arXiv:1303.3348 [hep-ph];
    H.~Dreiner, M.~Huck, M.~Kramer, D.~Schmeier and J.~Tattersall,
  %``Illuminating Dark Matter at the ILC,''
  Phys.\ Rev.\ D {\bf 87} (2013) 075015
  [arXiv:1211.2254 [hep-ph]];
  %%CITATION = ARXIV:1211.2254;%%
  %4 citations counted in INSPIRE as of 22 May 2013
  %%CITATION = ARXIV:1303.3348;%%
   J.~Kopp, E.~T.~Neil, R.~Primulando and J.~Zupan,
  %``From gamma ray line signals of dark matter to the LHC,''
  Phys.\  Dark.\  Univ.\  {\bf 2} (2013) 22
  [arXiv:1301.1683 [hep-ph]];
  %%CITATION = ARXIV:1301.1683;%%
  %4 citations counted in INSPIRE as of 22 May 2013
   M.~T.~Frandsen, F.~Kahlhoefer, A.~Preston, S.~Sarkar and K.~Schmidt-Hoberg,
  %``LHC and Tevatron Bounds on the Dark Matter Direct Detection Cross-Section for Vector Mediators,''
  JHEP {\bf 1207} (2012) 123
  [arXiv:1204.3839 [hep-ph]];
  %%CITATION = ARXIV:1204.3839;%%
  %12 citations counted in INSPIRE as of 22 May 2013
   J.~Goodman and W.~Shepherd,
  %``LHC Bounds on UV-Complete Models of Dark Matter,''
  arXiv:1111.2359 [hep-ph];
  %%CITATION = ARXIV:1111.2359;%%
  %10 citations counted in INSPIRE as of 22 May 2013
  P.~J.~Fox, R.~Harnik, J.~Kopp and Y.~Tsai,
  %``Missing Energy Signatures of Dark Matter at the LHC,''
  Phys.\ Rev.\ D {\bf 85} (2012) 056011
  [arXiv:1109.4398 [hep-ph]];
  %%CITATION = ARXIV:1109.4398;%%
  %77 citations counted in INSPIRE as of 22 May 2013
   Y.~Mambrini and B.~Zaldivar,
  %``When LEP and Tevatron combined with WMAP and XENON100 shed light on the nature of Dark Matter,''
  JCAP {\bf 1110} (2011) 023
  [arXiv:1106.4819 [hep-ph]].
  %%CITATION = ARXIV:1106.4819;%%
  %21 citations counted in INSPIRE as of 22 May 2013
  
  
  
  
  %==================================  INDIRECT  ==================================
  
  
  \bibitem{Cheung:2012gi}
  K.~Cheung, P.~-Y.~Tseng, Y.~-L.~S.~Tsai and T.~-C.~Yuan,
  %``Global Constraints on Effective Dark Matter Interactions: Relic Density, Direct Detection, Indirect Detection, and Collider,''
  JCAP {\bf 1205} (2012) 001
  [arXiv:1201.3402 [hep-ph]];
  %%CITATION = ARXIV:1201.3402;%%
  %22 citations counted in INSPIRE as of 22 May 2013
   C.~R.~Das, O.~Mena, S.~Palomares-Ruiz and S.~Pascoli,
  %``Determining the Dark Matter Mass with DeepCore,''
  arXiv:1110.5095 [hep-ph];
  %%CITATION = ARXIV:1110.5095;%%
  %4 citations counted in INSPIRE as of 22 May 2013
   C.~-L.~Shan,
  %``Effects of Residue Background Events in Direct Dark Matter Detection Experiments on the Estimation of the Spin-Independent WIMP-Nucleon Coupling,''
  arXiv:1103.4049 [hep-ph];
  %%CITATION = ARXIV:1103.4049;%%
  %3 citations counted in INSPIRE as of 22 May 2013
  M.~Pato, L.~Baudis, G.~Bertone, R.~Ruiz de Austri, L.~E.~Strigari and R.~Trotta,
  %``Complementarity of Dark Matter Direct Detection Targets,''
  Phys.\ Rev.\ D {\bf 83} (2011) 083505
  [arXiv:1012.3458 [astro-ph.CO]];
  %%CITATION = ARXIV:1012.3458;%%
  %37 citations counted in INSPIRE as of 22 May 2013
    G.~Bertone, D.~G.~Cerdeno, M.~Fornasa, R.~R.~de Austri and R.~Trotta,
  %``Identification of Dark Matter particles with LHC and direct detection data,''
  Phys.\ Rev.\ D {\bf 82} (2010) 055008
  [arXiv:1005.4280 [hep-ph]];
  %%CITATION = ARXIV:1005.4280;%%
  %36 citations counted in INSPIRE as of 22 May 2013
   N.~Bernal, A.~Goudelis, Y.~Mambrini and C.~Munoz,
  %``Determining the WIMP mass using the complementarity between direct and indirect searches and the ILC,''
  JCAP {\bf 0901} (2009) 046
  [arXiv:0804.1976 [hep-ph]];
  %%CITATION = ARXIV:0804.1976;%%
  %36 citations counted in INSPIRE as of 22 May 2013
    O.~Mena, S.~Palomares-Ruiz and S.~Pascoli,
  %``Reconstructing WIMP properties with neutrino detectors,''
  Phys.\ Lett.\ B {\bf 664} (2008) 92
  [arXiv:0706.3909 [hep-ph]];
  %%CITATION = ARXIV:0706.3909;%%
  %21 citations counted in INSPIRE as of 22 May 2013
   S.~Palomares-Ruiz and J.~M.~Siegal-Gaskins,
  %``Annihilation vs. Decay: Constraining dark matter properties from a gamma-ray detection,''
  JCAP {\bf 1007} (2010) 023
  [arXiv:1003.1142 [astro-ph.CO]];
  %%CITATION = ARXIV:1003.1142;%%
  %18 citations counted in INSPIRE as of 22 May 2013
  P.~Konar, K.~Kong, K.~T.~Matchev and M.~Perelstein,
  %``Shedding Light on the Dark Sector with Direct WIMP Production,''
  New J.\ Phys.\  {\bf 11} (2009) 105004
  [arXiv:0902.2000 [hep-ph]].
  %%CITATION = ARXIV:0902.2000;%%
  %21 citations counted in INSPIRE as of 22 May 2013
  
  
   %======================  FIMP  =========================================


   \bibitem{Moroi:1993mb}
  T.~Moroi, H.~Murayama and M.~Yamaguchi,
  %``Cosmological constraints on the light stable gravitino,''
  Phys.\ Lett.\ B {\bf 303} (1993) 289.
  %%CITATION = PHLTA,B303,289;%%
  %356 citations counted in INSPIRE as of 23 May 2013

  
  
  %\cite{Hall:2009bx}
\bibitem{Hall:2009bx}
  L.~J.~Hall, K.~Jedamzik, J.~March-Russell and S.~M.~West,
 % ``Freeze-In Production of FIMP Dark Matter,''
  JHEP {\bf 1003} (2010) 080
  [arXiv:0911.1120 [hep-ph]].
  %%CITATION = ARXIV:0911.1120;%%
  %45 citations counted in INSPIRE as of 17 Feb 2013
%\cite{McDonald:2001vt}
  
  \bibitem{Yaguna:2011ei}
   C.~E.~Yaguna,
  %``An intermediate framework between WIMP, FIMP, and EWIP dark matter,''
  JCAP {\bf 1202} (2012) 006
  [arXiv:1111.6831 [hep-ph]].
  %%CITATION = ARXIV:1111.6831;%%
  %2 citations counted in INSPIRE as of 18 Jun 2013
  
  
  %\cite{Chu:2011be}
\bibitem{Chu:2011be}
  X.~Chu, T.~Hambye and M.~H.~G.~Tytgat,
  %``The Four Basic Ways of Creating Dark Matter Through a Portal,''
  JCAP {\bf 1205} (2012) 034
  [arXiv:1112.0493 [hep-ph]].

  \bibitem{McDonald:2001vt}
  J.~McDonald,
 % ``Thermally generated gauge singlet scalars as selfinteracting dark matter'',
  Phys.\ Rev.\ Lett.\  {\bf 88} (2002) 091304
  [hep-ph/0106249].
  %%CITATION = HEP-PH/0106249;%%
  %59 citations counted in INSPIRE as of 17 Feb 2013
  
\bibitem{Yaguna:2011qn}
  C.~E.~Yaguna,
  %``The Singlet Scalar as FIMP Dark Matter,''
  JHEP {\bf 1108} (2011) 060
  [arXiv:1105.1654 [hep-ph]].
  %%CITATION = ARXIV:1105.1654;%%
  %4 citations counted in INSPIRE as of 04 May 2013


  
  
  \bibitem{Arcadi:2013aba}
  G.~Arcadi and L.~Covi,
  %``Minimal Decaying Dark Matter and the LHC,''
  arXiv:1305.6587 [hep-ph].
  %%CITATION = ARXIV:1305.6587;%%
  
  
  
%\cite{Mambrini:2013iaa}
\bibitem{Mambrini:2013iaa}
  Y.~Mambrini, K.~A.~Olive, J.~Quevillon and B.~Zaldivar,
  %``Gauge Coupling Unification and Non-Equilibrium Thermal Dark Matter,''
  arXiv:1302.4438 [hep-ph].
  %%CITATION = ARXIV:1302.4438;%%


%========================= EFFECTIVE COUPLINGS  ==================================


 %%%%%%%%%%%%%%  EFFECTIVE APPROACH, COLLIDER%%%%%%%%%%%%%%%%   
%\cite{Bai:2010hh}
\bibitem{Bai:2010hh}
  Y.~Bai, P.~J.~Fox and R.~Harnik,
  %``The Tevatron at the Frontier of Dark Matter Direct Detection,''
  JHEP {\bf 1012} (2010) 048
  [arXiv:1005.3797 [hep-ph]].
  %%CITATION = ARXIV:1005.3797;%%
  %99 citations counted in INSPIRE as of 17 Feb 2013
  %\cite{Fox:2011fx}
\bibitem{Fox:2011fx}
  P.~J.~Fox, R.~Harnik, J.~Kopp and Y.~Tsai,
  %``LEP Shines Light on Dark Matter,''
  Phys.\ Rev.\ D {\bf 84} (2011) 014028
  [arXiv:1103.0240 [hep-ph]].
  %%CITATION = ARXIV:1103.0240;%%
  %44 citations counted in INSPIRE as of 17 Feb 2013
  %\cite{Mambrini:2011pw}
\bibitem{Mambrini:2011pw}
  Y.~Mambrini and B.~Zaldivar,
  %``When LEP and Tevatron combined with WMAP and XENON100 shed light on the nature of Dark Matter,''
  JCAP {\bf 1110} (2011) 023
  [arXiv:1106.4819 [hep-ph]].
  %%CITATION = ARXIV:1106.4819;%%
  %19 citations counted in INSPIRE as of 17 Feb 2013   
   %\cite{Kamenik:2011nb}
\bibitem{Kamenik:2011nb}
  J.~F.~Kamenik and J.~Zupan,
  %``Discovering Dark Matter Through Flavor Violation at the LHC,''
  Phys.\ Rev.\ D {\bf 84} (2011) 111502
  [arXiv:1107.0623 [hep-ph]].
  %%CITATION = ARXIV:1107.0623;%%
  %13 citations counted in INSPIRE as of 17 Feb 2013
 %\cite{Fox:2011pm}
\bibitem{Fox:2011pm}
  P.~J.~Fox, R.~Harnik, J.~Kopp and Y.~Tsai,
  %``Missing Energy Signatures of Dark Matter at the LHC,''
  Phys.\ Rev.\ D {\bf 85} (2012) 056011
  [arXiv:1109.4398 [hep-ph]].
  %%CITATION = ARXIV:1109.4398;%%
  %67 citations counted in INSPIRE as of 17 Feb 2013
  %\cite{Goodman:2011jq}
\bibitem{Goodman:2011jq}
  J.~Goodman and W.~Shepherd,
  %``LHC Bounds on UV-Complete Models of Dark Matter,''
  arXiv:1111.2359 [hep-ph].
  %%CITATION = ARXIV:1111.2359;%%
  %9 citations counted in INSPIRE as of 17 Feb 2013
  %\cite{Frandsen:2012rk}
\bibitem{Frandsen:2012rk}
  M.~T.~Frandsen, F.~Kahlhoefer, A.~Preston, S.~Sarkar and K.~Schmidt-Hoberg,
  %``LHC and Tevatron Bounds on the Dark Matter Direct Detection Cross-Section for Vector Mediators,''
  JHEP {\bf 1207} (2012) 123
  [arXiv:1204.3839 [hep-ph]].
  %%CITATION = ARXIV:1204.3839;%%
  %7 citations counted in INSPIRE as of 17 Feb 2013
   %\cite{Dreiner:2012xm}

  %\cite{Chae:2012bq}
\bibitem{Chae:2012bq}
  Y.~J.~Chae and M.~Perelstein,
  %``Dark Matter Search at a Linear Collider: Effective Operator Approach,''
  arXiv:1211.4008 [hep-ph].
  %%CITATION = ARXIV:1211.4008;%%
  %1 citations counted in INSPIRE as of 17 Feb 2013
  
 %%%%%%%%%%%%%%  EFFECTIVE APPROACH, INDIRECT DETECTION %%%%%%%%%%%%%%%%
 %\cite{Gao:2011ka}
\bibitem{Gao:2011ka}
  X.~Gao, Z.~Kang and T.~Li,
  %``Origins of the Isospin Violation of Dark Matter Interactions,''
  JCAP {\bf 1301} (2013) 021
  [arXiv:1107.3529 [hep-ph]].
  %%CITATION = ARXIV:1107.3529;%%
  %19 citations counted in INSPIRE as of 17 Feb 2013 
  %\cite{Rajaraman:2012db}
\bibitem{Rajaraman:2012db}
  A.~Rajaraman, T.~M.~P.~Tait and D.~Whiteson,
  %``Two Lines or Not Two Lines? That is the Question of Gamma Ray Spectra,''
  JCAP {\bf 1209} (2012) 003
  [arXiv:1205.4723 [hep-ph]].
  %%CITATION = ARXIV:1205.4723;%%
  %\cite{Shoemaker:2011vi}
\bibitem{Shoemaker:2011vi}
  I.~M.~Shoemaker and L.~Vecchi,
  %``Unitarity and Monojet Bounds on Models for DAMA, CoGeNT, and CRESST-II,''
  Phys.\ Rev.\ D {\bf 86} (2012) 015023
  [arXiv:1112.5457 [hep-ph]].
  %%CITATION = ARXIV:1112.5457;%%
  %\cite{Kopp:2009et}
\bibitem{Kopp:2009et}
  J.~Kopp, V.~Niro, T.~Schwetz and J.~Zupan,
  %``DAMA/LIBRA and leptonically interacting Dark Matter,''
  Phys.\ Rev.\ D {\bf 80} (2009) 083502
  [arXiv:0907.3159 [hep-ph]].
  %\cite{Kopp:2010su}
\bibitem{Kopp:2010su}
  J.~Kopp, V.~Niro, T.~Schwetz and J.~Zupan,
  %``Leptophilic Dark Matter in Direct Detection Experiments and in the Sun,''
  PoS IDM {\bf 2010} (2011) 118
  [arXiv:1011.1398 [hep-ph]].
    %\cite{Mambrini:2012ue}
\bibitem{Mambrini:2012ue}
  Y.~Mambrini, M.~H.~G.~Tytgat, G.~Zaharijas and B.~Zaldivar,
  %``Complementarity of Galactic radio and collider data in constraining WIMP dark matter models,''
  JCAP {\bf 1211} (2012) 038
  [arXiv:1206.2352 [hep-ph]].
  %\cite{Gondolo:2012vh}
\bibitem{Gondolo:2012vh}
  P.~Gondolo, J.~Hisano and K.~Kadota,
  %``The Effect of quark interactions on dark matter kinetic decoupling and the mass of the smallest dark halos,''
  Phys.\ Rev.\ D {\bf 86} (2012) 083523
  [arXiv:1205.1914 [hep-ph]].
  
 %======================================  Z'  =========================================

 \bibitem{Langacker:2008yv}
 P.~Langacker,
 %``The Physics of Heavy $Z^\prime$ Gauge Bosons,''
 Rev.\ Mod.\ Phys.\  {\bf 81} (2008) 1199
 [arXiv:0801.1345 [hep-ph]].





\bibitem{Holdom}
 R.~Foot, X.~-G.~He,
 %``Comment on Z Z-prime mixing in extended gauge theories,''
 Phys.\ Lett.\  {\bf B267 } (1991)  509-512;
   R.~Foot, H.~Lew, R.~R.~Volkas,
 % ``A Model with fundamental improper space-time symmetries,''
 Phys.\ Lett.\  {\bf B272 } (1991)  67-70;
 B.~Holdom,
 % ``Two U(1)'S And Epsilon Charge Shifts,''
 Phys.\ Lett.\  B {\bf 166}, 196 (1986);
 %%CITATION = PHLTA,B166,196;%%
 D.~Feldman, Z.~Liu and P.~Nath,
 %``The Stueckelberg Z' extension with kinetic mixing and milli-charged dark
 %matter from the hidden sector,''
 Phys.\ Rev.\  D {\bf 75} (2007) 115001
 [arXiv:hep-ph/0702123];
 %%CITATION = PHRVA,D75,115001;%%
 S.~P.~Martin,
 %``Implications of supersymmetric models with natural R-parity
%conservation,''
 Phys.\ Rev.\  D {\bf 54} (1996) 2340
 [arXiv:hep-ph/9602349];
 %%CITATION = PHRVA,D54,2340;%%
 T.~G.~Rizzo,
 %``Gauge kinetic mixing and leptophobic $Z^\prime$ in E(6) and SO(10),''
 Phys.\ Rev.\  D {\bf 59} (1999) 015020
 [arXiv:hep-ph/9806397];
 %%CITATION = PHRVA,D59,015020;%%
 F.~del Aguila, M.~Masip and M.~Perez-Victoria,
 %``Physical parameters and renormalization of U(1)-a x U(1)-b models,''
 Nucl.\ Phys.\  B {\bf 456} (1995) 531
 [arXiv:hep-ph/9507455];
 %%CITATION = NUPHA,B456,531;%%
 B.~A.~Dobrescu,
 %``Massless gauge bosons other than the photon,''
 Phys.\ Rev.\ Lett.\  {\bf 94} (2005) 151802
 [arXiv:hep-ph/0411004];
 %%CITATION = PRLTA,94,151802;%%
% K.~R.~Dienes, C.~F.~Kolda and J.~March-Russell,
 %``Kinetic mixing and the supersymmetric gauge hierarchy,''
% Nucl.\ Phys.\  B {\bf 492} (1997) 104
% [arXiv:hep-ph/9610479];
 %%CITATION = NUPHA,B492,104;%%
 T.~Cohen, D.~J.~Phalen, A.~Pierce and K.~M.~Zurek,
 %``Asymmetric Dark Matter from a GeV Hidden Sector,''
 arXiv:1005.1655 [hep-ph];
 %%CITATION = ARXIV:1005.1655;%%
 Z.~Kang, T.~Li, T.~Liu, C.~Tong, J.~M.~Yang,
 %``Light Dark Matter from the $U(1)_X$ Sector in the NMSSM with Gauge
%Mediation,''
 JCAP {\bf 1101 } (2011)  028.
 [arXiv:1008.5243 [hep-ph]].

\bibitem{Dienes:1996zr}
  K.~R.~Dienes, C.~F.~Kolda and J.~March-Russell,
  %``Kinetic mixing and the supersymmetric gauge hierarchy,''
  Nucl.\ Phys.\ B {\bf 492} (1997) 104
  [hep-ph/9610479].

\bibitem{Feldman:2006wd}
 D.~Feldman, B.~Kors and P.~Nath,
 %``Extra-weakly Interacting Dark Matter,''
 Phys.\ Rev.\  D {\bf 75}, 023503 (2007)
 [arXiv:hep-ph/0610133].
 %%CITATION = PHRVA,D75,023503;%%


%\cite{Cicoli:2011yh}
\bibitem{Cicoli:2011yh}
  M.~Cicoli, M.~Goodsell, J.~Jaeckel and A.~Ringwald,
%  ``Testing String Vacua in the Lab: From a Hidden CMB to Dark Forces in Flux
 % Compactifications,''
  arXiv:1103.3705 [hep-th].
  %%CITATION = ARXIV:1103.3705;%%

\bibitem{Kumar:2007zza}
  J.~Kumar, A.~Rajaraman and J.~D.~Wells,
 % ``Probing the Green-Schwarz Mechanism at the Large Hadron Collider,''
  Phys.\ Rev.\  D {\bf 77} (2008) 066011
  [arXiv:0707.3488 [hep-ph]].
  %%CITATION = PHRVA,D77,066011;%%

 
 \bibitem{Goodsell:2011wn}
  M.~Goodsell, S.~Ramos-Sanchez and A.~Ringwald,
  %``Kinetic Mixing of U(1)s in Heterotic Orbifolds,''
  JHEP {\bf 1201} (2012) 021
  [arXiv:1110.6901 [hep-th]].
  %%CITATION = ARXIV:1110.6901;%%
  %10 citations counted in INSPIRE as of 12 Jun 2013

 
 
\bibitem{Javier}
 M.~Goodsell, J.~Jaeckel, J.~Redondo and A.~Ringwald,
 % ``Naturally Light Hidden Photons in LARGE Volume String Compactifications,''
  JHEP {\bf 0911} (2009) 027
  [arXiv:0909.0515 [hep-ph]];
   %%CITATION = JHEPA,0911,027;%%
 S.~A.~Abel, M.~D.~Goodsell, J.~Jaeckel, V.~V.~Khoze and A.~Ringwald,
 % ``Kinetic Mixing of the Photon with Hidden U(1)s in String Phenomenology,''
  JHEP {\bf 0807} (2008) 124
  [arXiv:0803.1449 [hep-ph]];
  %%CITATION = JHEPA,0807,124;%%



\bibitem{Cassel:2009pu}
  S.~Cassel, D.~M.~Ghilencea and G.~G.~Ross,
%  ``Electroweak and Dark Matter Constraints on a Z' in Models with a Hidden
 % Valley,''
  Nucl.\ Phys.\  B {\bf 827} (2010) 256
  [arXiv:0903.1118 [hep-ph]].
  %%CITATION = NUPHA,B827,256;%%

\bibitem{Andreas:2011in}
  S.~Andreas, M.~D.~Goodsell and A.~Ringwald,
  %``Dark Matter and Dark Forces from a Supersymmetric Hidden Sector,''
  Phys.\ Rev.\ D {\bf 87} (2013) 025007
  [arXiv:1109.2869 [hep-ph]].
  %%CITATION = ARXIV:1109.2869;%%
  %17 citations counted in INSPIRE as of 12 Jun 2013


\bibitem{Krauss:2013jva}
  M.~E.~Krauss, W.~Porod and F.~Staub,
  %``SO(10) inspired GMSB,''
  arXiv:1304.0769 [hep-ph].
  %%CITATION = ARXIV:1304.0769;%%



\bibitem{Feldman:2007wj}
 D.~Feldman, Z.~Liu and P.~Nath,
 %``The Stueckelberg Z-prime Extension with Kinetic Mixing and
%Milli-Charged Dark Matter From the Hidden Sector,''
 Phys.\ Rev.\ D {\bf 75} (2007) 115001
 [hep-ph/0702123 [HEP-PH]].
 %%CITATION = HEP-PH/0702123;%%


\bibitem{Pospelov:2008zw}
  M.~Pospelov,
  %``Secluded U(1) below the weak scale,''
  Phys.\ Rev.\ D {\bf 80} (2009) 095002
  [arXiv:0811.1030 [hep-ph]].
  %%CITATION = ARXIV:0811.1030;%%
  %148 citations counted in INSPIRE as of 29 May 2013




      %\cite{Mambrini:2010yp}
\bibitem{Mambrini:2010yp}
 Y.~Mambrini,
  %``The ZZ' kinetic mixing in the light of the recent direct and indirect dark matter searches,''
  JCAP {\bf 1107} (2011) 009
  [arXiv:1104.4799 [hep-ph]];
  %%CITATION = ARXIV:1104.4799;%%
  %29 citations counted in INSPIRE as of 29 May 2013
   Y.~Mambrini,
  %``The Kinetic dark-mixing in the light of CoGENT and XENON100,''
  JCAP {\bf 1009} (2010) 022
  [arXiv:1006.3318 [hep-ph]];
  %%CITATION = ARXIV:1006.3318;%%
  E.~J.~Chun, J.~C.~Park and S.~Scopel,
  ``Dark matter and a new gauge boson through kinetic mixing,''
  JHEP {\bf 1102} (2011) 100
  [arXiv:1011.3300 [hep-ph]].
  %%CITATION = JHEPA,1102,100;%%

\bibitem{Domingo:2013tna}
  F.~Domingo, O.~Lebedev, Y.~Mambrini, J.~Quevillon and A.~Ringwald,
  %``More on the Hypercharge Portal into the Dark Sector,''
  arXiv:1305.6815 [hep-ph];
  %%CITATION = ARXIV:1305.6815;%%
  %1 citations counted in INSPIRE as of 18 Jun 2013
    Y.~Farzan and A.~R.~Akbarieh,
  %``Natural explanation for 130 GeV photon line within vector boson dark matter model,''
  arXiv:1211.4685 [hep-ph].
  %%CITATION = ARXIV:1211.4685;%%
  %9 citations counted in INSPIRE as of 18 Jun 2013

\bibitem{Heeck:2011md}
  J.~Heeck and W.~Rodejohann,
  %``Kinetic and mass mixing with three abelian groups,''
  Phys.\ Lett.\ B {\bf 705} (2011) 369
  [arXiv:1109.1508 [hep-ph]].
  %%CITATION = ARXIV:1109.1508;%%
  %10 citations counted in INSPIRE as of 18 Jun 2013

\bibitem{monogamma}
 E.~Dudas, Y.~Mambrini, S.~Pokorski and A.~Romagnoni,
  %``Extra U(1) as natural source of a monochromatic gamma ray line,''
  JHEP {\bf 1210} (2012) 123
  [arXiv:1205.1520 [hep-ph]];
  %%CITATION = ARXIV:1205.1520;%%
  %56 citations counted in INSPIRE as of 29 May 2013
  Y.~Mambrini,
  %``A clear Dark Matter gamma ray line generated by the Green-Schwarz
  mechanism,''
  JCAP {\bf 0912}, 005 (2009)
  [arXiv:0907.2918 [hep-ph]];
  %%CITATION = JCAPA,0912,005;%%
    E.~Dudas, Y.~Mambrini, S.~Pokorski and A.~Romagnoni,
 % ``(In)visible Z' and dark matter,''
  JHEP {\bf 0908}, 014 (2009)
  [arXiv:0904.1745 [hep-ph]];
    C.~B.~Jackson, G.~Servant, G.~Shaughnessy, T.~M.~P.~Tait and M.~Taoso,
  %``Gamma-ray lines and One-Loop Continuum from s-channel Dark Matter Annihilations,''
  arXiv:1302.1802 [hep-ph];
  %%CITATION = ARXIV:1302.1802;%%
  %8 citations counted in INSPIRE as of 18 Jun 2013
  %%CITATION = JHEPA,0908,014;%%
  C.~B.~Jackson, G.~Servant, G.~Shaughnessy, T.~M.~P.~Tait and M.~Taoso,
  %``Higgs in Space!,''
  JCAP {\bf 1004}, 004 (2010)
  [arXiv:0912.0004 [hep-ph]].
  %%CITATION = JCAPA,1004,004;%%



\bibitem{Baumgart:2009tn}
  M.~Baumgart, C.~Cheung, J.~T.~Ruderman, L.~T.~Wang and I.~Yavin,
  %``Non-Abelian Dark Sectors and Their Collider Signatures,''
  JHEP {\bf 0904} (2009) 014
  [arXiv:0901.0283 [hep-ph]].
  %%CITATION = JHEPA,0904,014;%%

%===================================   Effective analysis  ==============================

%===================================================================================


\bibitem{bfmz} 
  M.~Blennow, E.~Fernandez-Martinez and B.~Zaldivar,
  %``Freeze-in through portals,''
  arXiv:1309.7348 [hep-ph].
  %%CITATION = ARXIV:1309.7348;%%


\bibitem{Arcadi:2013qia}
  G.~Arcadi, Y.~Mambrini, M.~H.~G.~Tytgat and B.~Zaldivar,
  %``Invisible Z' and dark matter: LHC vs LUX constraints,''
  arXiv:1401.0221 [hep-ph].
  %%CITATION = ARXIV:1401.0221;%%
  %3 citations counted in INSPIRE as of 27 Jan 2014


%=========================  WMAP CONSTRAINT  ========================================

  
  




%==================================  DIRECT DETECTION    ================================


\bibitem{Aprile:2011hi}
  E.~Aprile {\it et al.}  [XENON100 Collaboration],
  %``Dark Matter Results from 100 Live Days of XENON100 Data,''
  Phys.\ Rev.\ Lett.\  {\bf 107} (2011) 131302
  [arXiv:1104.2549 [astro-ph.CO]];
  %%CITATION = ARXIV:1104.2549;%%
  %494 citations counted in INSPIRE as of 05 Jun 2013
  E.~Aprile {\it et al.}  [XENON100 Collaboration],
  %``First Dark Matter Results from the XENON100 Experiment,''
  Phys.\ Rev.\ Lett.\  {\bf 105} (2010) 131302
  [arXiv:1005.0380 [astro-ph.CO]];
  %%CITATION = ARXIV:1005.0380;%%
  %337 citations counted in INSPIRE as of 05 Jun 2013
 E.~Aprile {\it et al.}  [XENON100 Collaboration],
  %``Dark Matter Results from 225 Live Days of XENON100 Data,''
  Phys.\ Rev.\ Lett.\  {\bf 109} (2012) 181301
  [arXiv:1207.5988 [astro-ph.CO]].
  %%CITATION = ARXIV:1207.5988;%%
  %242 citations counted in INSPIRE as of 05 Jun 2013

%\bibitem{Mambrini:2012ue}
%  Y.~Mambrini, M.~H.~G.~Tytgat, G.~Zaharijas and B.~Zaldivar,
%  %``Complementarity of Galactic radio and collider data in constraining WIMP dark matter models,''
%  JCAP {\bf 1211} (2012) 038
%  [arXiv:1206.2352 [hep-ph]].
%  %%CITATION = ARXIV:1206.2352;%%
%  %4 citations counted in INSPIRE as of 05 Jun 2013



\end{thebibliography}
\end{document}